\documentclass[twocolumn]{jpsj3} 

\usepackage{graphicx, color}
\usepackage{subfigure}

\def\nle{\ \raise.3ex\hbox{$<$}\kern-0.8em\lower.7ex\hbox{$\sim$}\ }
\def\nge{\ \raise.3ex\hbox{$>$}\kern-0.8em\lower.7ex\hbox{$\sim$}\ }

\title{Spin-Wave Theory for the Scalar Chiral Phase in the Multiple-Spin Exchange Model on a Triangular Lattice}

\author{Sh\^{o}go Taira$^{\rm 1}$, Chitoshi Yasuda$^{\rm 2,}$\thanks{E-mail address: cyasuda@sci.u-ryukyu.ac.jp}, Tsutomu Momoi$^{\rm 3,4}$, and Kenn Kubo$^{\rm 5}$}

\inst{$^{\rm 1}$
Graduate School of Engineering and Science, University of the Ryukyus, Nishihara, Okinawa 903-0123, Japan\\
$^{\rm 2}$
Department of Physics and Earth Sciences, University of the Ryukyus, Nishihara, Okinawa 903-0213, Japan\\
$^{\rm 3}$
Condensed Matter Theory Laboratory, RIKEN, Wako, Saitama 351-0198, Japan\\
$^{\rm 4}$
RIKEN Center for Emergent Matter Science (CEMS), Wako, Saitama 351-0198, Japan\\
$^{\rm 5}$
Department of Physics and Mathematics, Aoyama Gakuin University, Sagamihara 252-5258, Japan}

\abst{
We study the effects of quantum fluctuations on a non-coplanar tetrahedral spin structure, which has a scalar chiral order, in the spin-1/2 multiple-spin exchange model with up to the six-spin exchange interactions on a triangular lattice.
We find that, in the linear spin-wave approximation, the tetrahedral structure survives the quantum fluctuations because spin waves do not soften in the whole parameter region of the tetrahedral-structure phase evaluated for the classical system.
In the quantum corrections to the ground-state energy, sublattice magnetization, and scalar chirality, the effects of the quantum fluctuations are small for the ferromagnetic nearest-neighbor interactions and for the strong five-spin interactions.
The six-spin interactions have little effect on the quantum corrections in the tetrahedral-structure phase.
This calculation also corrects an error in the previously reported value of scalar chirality for the spin-1/2 multiple-spin exchange model with up to the four-spin exchange interactions.
}

\recdate{\today}

\begin{document}
\maketitle


\section{Introduction}

The solid phases of $^3$He layers are attracting extensive interest as a frustrated quantum spin system~\cite{Fukuyama2008}.
In a solid $^{3}$He thin film adsorbed on graphite, a solid layer with a triangular-lattice structure of $S = 1 / 2$ nuclear spins  is formed on the second layer~\cite{Franco1986, Godfrin1988, Greywall1990, Siqueira1996, Ishida1997}.
In the exchange process of the $^3$He atoms, there are not only exchange interactions between two neighboring atoms but also those among three or more neighboring atoms~\cite{Roger1983}.
The exchange integrals of three- and four-spin cyclic exchange interactions are larger than that of the two-spin exchange interaction owing to stoichiometric hindrance.
The second layer of the solid $^{3}$He thin film can be transformed from ferromagnets to highly frustrated antiferromagnets by tuning the coverage density from high to low.
By changing the density, the ratios of exchange couplings markedly shift.
Thus, the multiple-spin exchange (MSE) model on the triangular lattice has been extensively studied in various parameter regimes in relation to two-dimensional solid $^3$He.
From a comparison of experimental and theoretical results~\cite{Roger1990, Bernu1992,  Siqueira1996, Misguich1998, Misguich1999, Masutomi2004}, it is known that the five- and six-spin exchange interactions are not negligible for two-dimensional solid $^{3}$He.
Furthermore, exact-diagonalization studies of the MSE model with up to the six-spin exchange interactions have suggested that a nonmagnetic gapped spin-liquid phase is realized~\cite{Misguich1998, Misguich1999}.
The region of the spin-liquid phase overlaps with that of the tetrahedral-structure phase obtained by the mean-field approximation~\cite{Kubo1997}.

The MSE model on a triangular lattice also attracts interest as an effective model of organic triangular-lattice systems, $\kappa$-(BEDT-TTF)$_{2}^{}X$ and $Y$[Pd(demit)$_{2}^{}$]$_{2}^{}$.
In organic triangular-lattice systems, both long-range magnetic ordered and spin-liquid phases have been experimentally observed~\cite{Motrunich2005, Holt2014}.

Furthermore, the effects of quantum spin fluctuations on the scalar chiral ordering, which is realized in the tetrahedral-structure state of the classical ground state in the MSE model, have attracted attention.
It is known that quantum fluctuations have weak effects on the ordered phase with the chirality formed by an adjacent spin structure~\cite{Momoi1992}.
It is interesting to study whether the chiral order obtained within the mean-field approximation can survive the quantum fluctuations.
The scalar chiral ordering of the tetrahedral structure is also interesting in relation to peculiar transport properties in itinerant systems~\cite{Akagi2013}.

The aim of this paper is to elucidate the effects of quantum fluctuations on the tetrahedral spin structure with scalar chirality in the MSE model with up to the six-spin exchange interactions.
Focusing on the tetrahedral-structure state, we investigate the stability of the ground state against quantum fluctuations and the effects of the MSE interactions on the classical structure using linear spin-wave theory.
Studies of the MSE model with up to the four-spin exchange interactions using spin-wave theory have already been reported assuming the 120$^{\circ}$-structure,~\cite{Kubo1998, Yasuda2006} uuud~\cite{Kubo1998, Momoi1999}, and tetrahedral-structure~\cite{Momoi1997, Kubo1998-2} states as the ground state.

The remainder of this article is organized as follows.
We introduce the MSE model Hamiltonian with up to the six-spin exchange interactions in Sect.~2 and explain the classical phase diagram in Sect.~3.
We calculate the wave-number dependences of the spin-wave spectra using linear spin-wave theory in Sect.~4.
Furthermore, in Sect. 5, we examine the ground-state energy, sublattice magnetization, and scalar chirality for two systems, without second- and third-nearest-neighbor interactions and without five- and six-spin exchange interactions.
Finally, we devote Sect.~6 to a summary and discussion.


\section{MSE Model on a Triangular Lattice}

The Hamiltonian of the spin-1/2 MSE model with up to the six-spin exchange interactions on a triangular lattice is given by
\begin{align}
\mathcal{H}
    	= \sum_{n = 2}^{6} \mathcal{H}_{n}^{}
\label{eq_H}
\end{align}
with
\begin{align}
\mathcal{H}_{n}^{}
	= (-1)^{n} J_{n}^{} \sum_{n\mbox{-}{\rm spin~ring}} (P_{n}^{} + P_{n}^{-1})~,
	\label{H_n}
\end{align}
where $J_{n}^{}$ represents the positive exchange constants of $n$-spin ring exchanges for any $n$, and $P_{n}^{}$ and $P_{n}^{-1}$ are the $n$-spin ring exchange operators and their inverse operators, respectively~\cite{Thouless1965}.

The two- and three-spin Hamiltonians are written as
\begin{align}
\mathcal{H}_{2}^{}
	= \frac{J_{2}^{}}{2} \sum_{\rm bond} (1 + \mib{\sigma}_{1}^{} \cdot \mib{\sigma}_{2}^{})
	\label{H2}
\end{align}
and
\begin{align}
\mathcal{H}_{3}^{}
	= -\frac{J_{3}^{}}{2} \sum_{\rm triangle}
	(1 + \mib{\sigma}_{1}^{} \cdot \mib{\sigma}_{2}^{}
	+ \mib{\sigma}_{2}^{} \cdot \mib{\sigma}_{3}^{}
	+ \mib{\sigma}_{3}^{} \cdot \mib{\sigma}_{1}^{})~,
	\label{H3}
\end{align}
where $\sum_{\rm bond}$ and $\sum_{\rm triangle}$ are the summations taken over all bonds and all smallest triangles on the triangular lattice, respectively.
The subscripts of the Pauli spin operator $\mib{\sigma}_{i}^{}$ denote the sites belonging to each bond or each triangle.
Hereafter, the $\mib{\sigma}_{i}^{}$ is called spin.
The four-spin Hamiltonian $\mathcal{H}_{4}^{}$ is written as
\begin{align}
\mathcal{H}_{4}^{}
	&= \frac{J_{4}^{}}{4} \sum_{\rm plaq} \Bigg \{
	1 + \sum_{1 \leq \alpha < \beta \leq 4} \mib{\sigma}_{\alpha}^{} \cdot \mib{\sigma}_{\beta}^{}
	+ (\mib{\sigma}_{1}^{} \cdot \mib{\sigma}_{2}^{})(\mib{\sigma}_{3}^{} \cdot \mib{\sigma}_{4}^{})
	\notag \\
	&
	~~~~~
		+ (\mib{\sigma}_{1}^{} \cdot \mib{\sigma}_{4}^{})(\mib{\sigma}_{2}^{} \cdot \mib{\sigma}_{3}^{})
	- (\mib{\sigma}_{1}^{} \cdot \mib{\sigma}_{3}^{})(\mib{\sigma}_{2}^{} \cdot \mib{\sigma}_{4}^{}) \Bigg \}~,
	\label{H4}
\end{align}
where $\sum_{\rm plaq}$ is the summation taken over all the smallest diamonds.
The subscripts of $\mib{\sigma}_{i}^{}$ denote the sites belonging to each diamond, where (1, 3) and (2, 4) are diagonal bonds of the diamond.
The five-spin Hamiltonian $\mathcal{H}_{5}^{}$ is written as
\begin{align}
\mathcal{H}_{5}^{}
	&= -\frac{J_{5}^{}}{8} \sum_{\rm trap} \Bigg[
	1 + \sum_{1 \leq \alpha < \beta \leq 5} \mib{\sigma}_{\alpha}^{} \cdot \mib{\sigma}_{\beta}^{}
	\notag \\
	&~~
	+ \sum_{l = 1}^{5}
	\{ (\mib{\sigma}_{\alpha_{l}^{}}^{} \cdot \mib{\sigma}_{\beta_{l}^{}}^{})(\mib{\sigma}_{\gamma_{l}^{}}^{} \cdot \mib{\sigma}_{\delta_{l}^{}}^{})
	+ (\mib{\sigma}_{\alpha_{l}^{}}^{} \cdot \mib{\sigma}_{\delta_{l}^{}}^{})(\mib{\sigma}_{\beta_{l}^{}}^{} \cdot \mib{\sigma}_{\gamma_{l}^{}}^{})
		\notag \\
	&~~
	- (\mib{\sigma}_{\alpha_{l}^{}}^{} \cdot \mib{\sigma}_{\gamma_{l}^{}}^{})(\mib{\sigma}_{\beta_{l}^{}}^{} \cdot \mib{\sigma}_{\delta_{l}^{}}^{}) \} \Bigg]~,
	\label{H5}
\end{align}
where $\sum_{\rm trap}$ is the summation taken over all the smallest trapezoids.
The summation $\sum_{l = 1}^{5}$ is taken over all combinations of the four spins in each trapezoid.
The subscripts of $\mib{\sigma}_{i}^{}$ denote the sites belonging to a trapezoid, and
$\alpha_{l}^{} = l$,
$\beta_{l}^{} = {\rm mod}(l,5) + 1$,
$\gamma_{l}^{} = {\rm mod}(l + 1,5) + 1$,
$\delta_{l}^{} = {\rm mod}(l + 2,5) + 1$.
The six-spin Hamiltonian $\mathcal{H}_{6}^{}$ is written as
\begin{align}
\mathcal{H}_{6}^{}
	&= \frac{J_{6}^{}}{16} \sum_{\rm hexa} \Bigg[
	1 + \sum_{1 \leq \alpha < \beta \leq 6} \mib{\sigma}_{\alpha}^{} \cdot \mib{\sigma}_{\beta}^{}
	\notag \\
	&~~
	+ \sum_{l = 1}^{6}
	\{ (\mib{\sigma}_{\alpha_{l}^{}}^{} \cdot \mib{\sigma}_{\beta_{l}^{}}^{})(\mib{\sigma}_{\gamma_{l}^{}}^{} \cdot \mib{\sigma}_{\delta_{l}^{}}^{})
	+ (\mib{\sigma}_{\alpha_{l}^{}}^{} \cdot \mib{\sigma}_{\delta_{l}^{}}^{})(\mib{\sigma}_{\beta_{l}^{}}^{} \cdot \mib{\sigma}_{\gamma_{l}^{}}^{})
	\notag \\
	&~~
	- (\mib{\sigma}_{\alpha_{l}^{}}^{} \cdot \mib{\sigma}_{\gamma_{l}^{}}^{})(\mib{\sigma}_{\beta_{l}^{}}^{} \cdot \mib{\sigma}_{\delta_{l}^{}}^{}) \}
	\notag \\
	&~~
	+ \sum_{l = 1}^{6}
	\{ (\mib{\sigma}_{\alpha_{l}^{}}^{} \cdot \mib{\sigma}_{\beta_{l}^{}}^{})(\mib{\sigma}_{\gamma_{l}^{}}^{} \cdot \mib{\sigma}_{\zeta_{l}^{}}^{})
	+ (\mib{\sigma}_{\alpha_{l}^{}}^{} \cdot \mib{\sigma}_{\zeta_{l}^{}}^{})(\mib{\sigma}_{\beta_{l}^{}}^{} \cdot \mib{\sigma}_{\gamma_{l}^{}}^{})
	\notag \\
	&~~
	- (\mib{\sigma}_{\alpha_{l}^{}}^{} \cdot \mib{\sigma}_{\gamma_{l}^{}}^{})(\mib{\sigma}_{\beta_{l}^{}}^{} \cdot \mib{\sigma}_{\zeta_{l}^{}}^{}) \}
	\notag \\
	&~~
	+ \sum_{l = 1}^{3}
	\{ (\mib{\sigma}_{\alpha_{l}^{}}^{} \cdot \mib{\sigma}_{\beta_{l}^{}}^{})(\mib{\sigma}_{\delta_{l}^{}}^{} \cdot \mib{\sigma}_{\zeta_{l}^{}}^{})
	+ (\mib{\sigma}_{\alpha_{l}^{}}^{} \cdot \mib{\sigma}_{\zeta_{l}^{}}^{})(\mib{\sigma}_{\beta_{l}^{}}^{} \cdot \mib{\sigma}_{\delta_{l}^{}}^{})
	\notag \\
	&~~
	- (\mib{\sigma}_{\alpha_{l}^{}}^{} \cdot \mib{\sigma}_{\delta_{l}^{}}^{})(\mib{\sigma}_{\beta_{l}^{}}^{} \cdot \mib{\sigma}_{\zeta_{l}^{}}^{}) \} 
	\notag \\
	&~~
	+ \sum_{l = 1}^{2}
	(\mib{\sigma}_{\alpha_{l}^{}}^{} \cdot \mib{\sigma}_{\beta_{l}^{}}^{})
	(\mib{\sigma}_{\gamma_{l}^{}}^{} \cdot \mib{\sigma}_{\delta_{l}^{}}^{})
	(\mib{\sigma}_{\zeta_{l}^{}}^{} \cdot \mib{\sigma}_{\kappa_{l}^{}}^{})
	\notag \\
	&~~
	+ \sum_{l = 1}^{3}
	(\mib{\sigma}_{\alpha_{l}^{}}^{} \cdot \mib{\sigma}_{\delta_{l}^{}}^{})
	(\mib{\sigma}_{\beta_{l}^{}}^{} \cdot \mib{\sigma}_{\gamma_{l}^{}}^{})
	(\mib{\sigma}_{\zeta_{l}^{}}^{} \cdot \mib{\sigma}_{\kappa_{l}^{}}^{})
	\notag \\
	&~~
	+ \sum_{l = 1}^{3}
	(\mib{\sigma}_{\alpha_{l}^{}}^{} \cdot \mib{\sigma}_{\delta_{l}^{}}^{})
	(\mib{\sigma}_{\gamma_{l}^{}}^{} \cdot \mib{\sigma}_{\zeta_{l}^{}}^{})
	(\mib{\sigma}_{\beta_{l}^{}}^{} \cdot \mib{\sigma}_{\kappa_{l}^{}}^{})
	\notag \\
	&~~
	- \sum_{l = 1}^{6}
	(\mib{\sigma}_{\alpha_{l}^{}}^{} \cdot \mib{\sigma}_{\beta_{l}^{}}^{})
	(\mib{\sigma}_{\gamma_{l}^{}}^{} \cdot \mib{\sigma}_{\zeta_{l}^{}}^{})
	(\mib{\sigma}_{\delta_{l}^{}}^{} \cdot \mib{\sigma}_{\kappa_{l}^{}}^{})
	\notag \\
	&~~
	- (\mib{\sigma}_{1}^{} \cdot \mib{\sigma}_{4}^{})
	(\mib{\sigma}_{2}^{} \cdot \mib{\sigma}_{5}^{})
	(\mib{\sigma}_{3}^{} \cdot \mib{\sigma}_{6}^{}) \Bigg]~,
	\label{H6}
\end{align}
where $\sum_{\rm hexa}$ is the summation taken over all the smallest hexagons.
The summations of the products of four spins and six spins are taken over all combinations of the four spins and six spins in each hexagon, respectively.
The site indexes of $\mib{\sigma}_{i}^{}$ in a hexagon are defined as
$\alpha_{l}^{} = l$,
$\beta_{l}^{} = {\rm mod}(l,6) + 1$,
$\gamma_{l}^{} = {\rm mod}(l + 1,6) + 1$,
$\delta_{l}^{} = {\rm mod}(l + 2,6) + 1$,
$\zeta_{l}^{} = {\rm mod}(l + 3,6) + 1$,
$\kappa_{l}^{} = {\rm mod}(l + 4,6) + 1$.

Neglecting the constant terms and transforming the interactions as
\begin{align}
\mathcal{H}_{4}^{}
	= \frac{J_{4}^{}}{4} \sum_{\rm plaq} h_{4}^{}~,
	~~
\mathcal{H}_{5}^{}
	= -\frac{J_{5}^{}}{8} \sum_{\rm trap} h_{5}^{}~,
	~~
\mathcal{H}_{6}^{}
	= \frac{J_{6}^{}}{16} \sum_{\rm hexa} h_{6}^{}~,
	\label{H4-H6}
\end{align}
\begin{align}
J
	= \frac{J_{2}^{}}{2} - J_{3}^{}~,
	~~
K
	= \frac{J_{4}^{}}{4}~,
	~~
L
	= -\frac{J_{5}^{}}{8}~,
	~~
M
	= \frac{J_{6}^{}}{16}~,
	\label{J-M}
\end{align}
we obtain the Hamiltonian
\begin{align}
\mathcal{H}
	&= J\sum_{\rm n.n} \mib{\sigma}_{i}^{} \cdot \mib{\sigma}_{j}^{}
	+ K\sum_{\rm plaq} h_{4}^{}
	+ L\sum_{\rm trap} h_{5}^{}
	+ M\sum_{\rm hexa} h_{6}^{}~,
	\label{Ham_j2j3_0}
\end{align}
where $\sum_{\rm n.n}$ is the summation taken over all nearest-neighbor pairs.
We call the model the $J$-$K$-$L$-$M$ model.
$K$ and $M$ are always positive, and $L$ is always negative.
Because the magnitude of $J_{2}^{}$ is expected to be large compared with that of $J_{3}^{}$ at low atomic densities~\cite{Roger1983, Roger1990}, the value of $J$ is expected to be positive at low densities and negative at high densities in solid $^3$He.
In the present work, we take the value of $K$ to be a unit of energy.

In the present work, we also investigate the effects of the second- and third-nearest-neighbor interactions.
While these interactions are included in the five- and six-spin exchange interactions, their own effects on the system are interesting in relation to the effective model of organic triangular-lattice systems.
Thus, we investigate the systems described by
\begin{align}
\mathcal{H}
	&= J\sum_{\rm n.n} \mib{\sigma}_{i}^{} \cdot \mib{\sigma}_{j}^{}
	+ J_{\rm 2nd}^{} \sum_{\rm n.n.n} \mib{\sigma}_{i}^{} \cdot \mib{\sigma}_{j}^{}
	+ J_{\rm 3rd}^{} \sum_{\rm n.n.n.n} \mib{\sigma}_{i}^{} \cdot \mib{\sigma}_{j}^{}
	\notag \\
	&~~
	+ K\sum_{\rm plaq} h_{4}^{}
	+ L\sum_{\rm trap} h_{5}^{}
	+ M\sum_{\rm hexa} h_{6}^{}~,
	\label{Ham_J-M}
\end{align}
where $J_{\rm 2nd}^{}$ and $J_{\rm 3rd}^{}$ are the exchange integrals of the second- and third-nearest-neighbor interactions, and $\sum_{\rm n.n.n}$ and $\sum_{\rm n.n.n.n}$ are the summations taken over all second- and third-nearest-neighbor pairs, respectively.
We call the model the $J$-$K$-$J_{\rm 2nd}^{}$-$J_{\rm 3rd}^{}$ model with the Hamiltonian in Eq.~(\ref{Ham_J-M}) for $L = M = 0$ and investigate the $J$-$K$-$J_{\rm 2nd}^{}$-$J_{\rm 3rd}^{}$ model in Sect.~5.2.


\section{Classical Ground States}

\begin{figure}[t]
	\centering
	\includegraphics[width=0.45\textwidth]{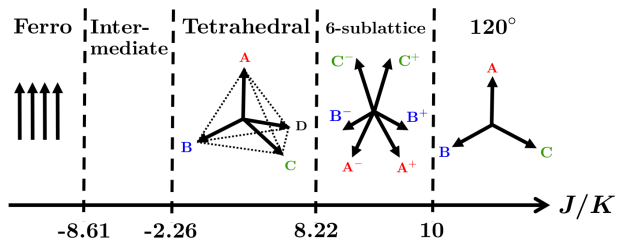}
	\caption{Phase diagram of the classical ground state parametrized by the two-spin exchange integral $J / K$ for $J_{\rm 2nd}^{} = J_{\rm 3rd}^{} = L = M = 0$.
	The tetrahedral and six-sublattice structures have a non-coplanar spin configuration, and the 120$^{\circ}$ structure has a coplanar spin configuration.}
	\label{j_phase_diagram}
\end{figure}
\begin{figure}[t]
	\centering
	\subfigure[]
	{\resizebox{0.35\textwidth}{!}{\includegraphics{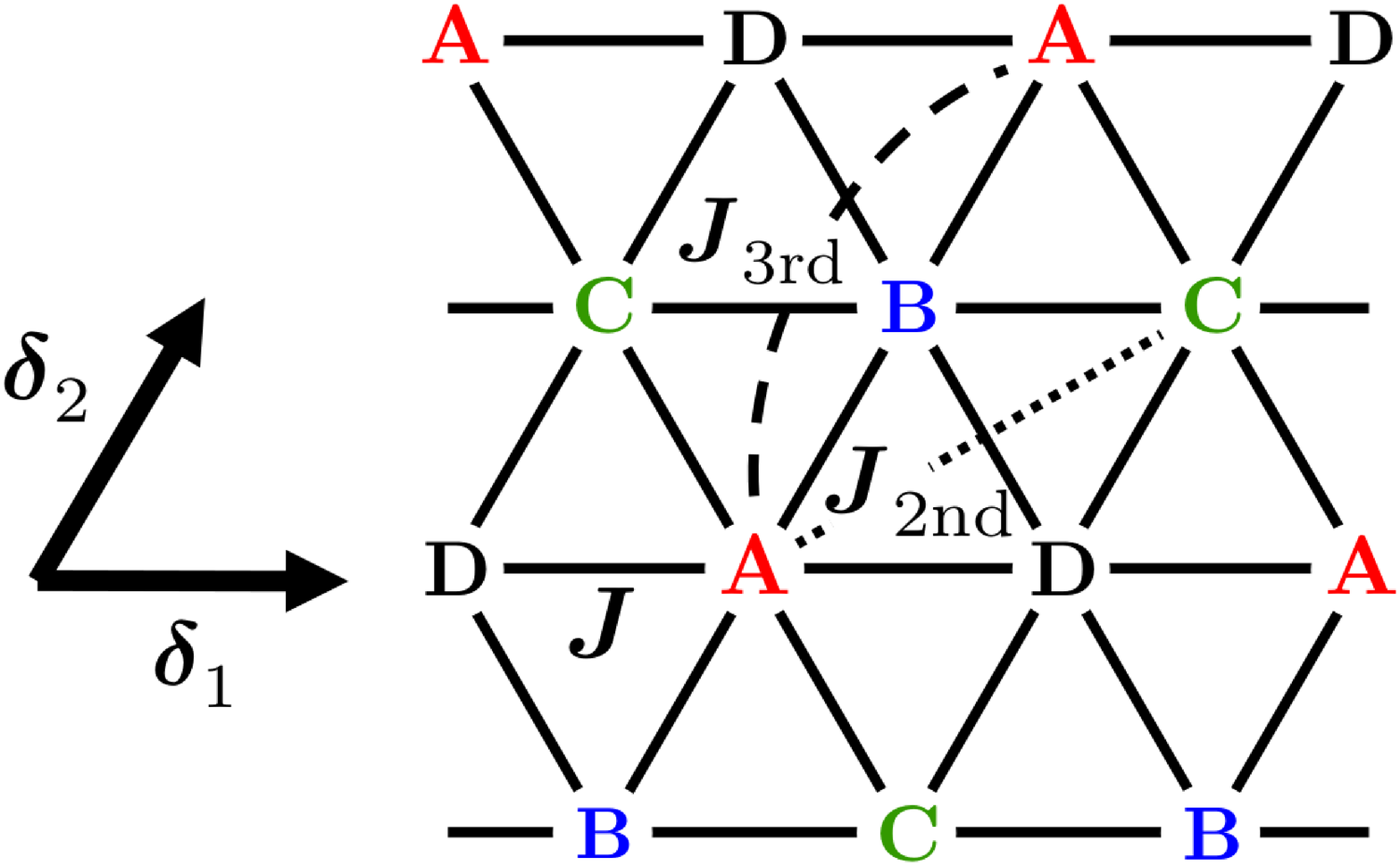}}}
	\hspace{2cm}
	\subfigure[]
	{\resizebox{0.25\textwidth}{!}{\includegraphics{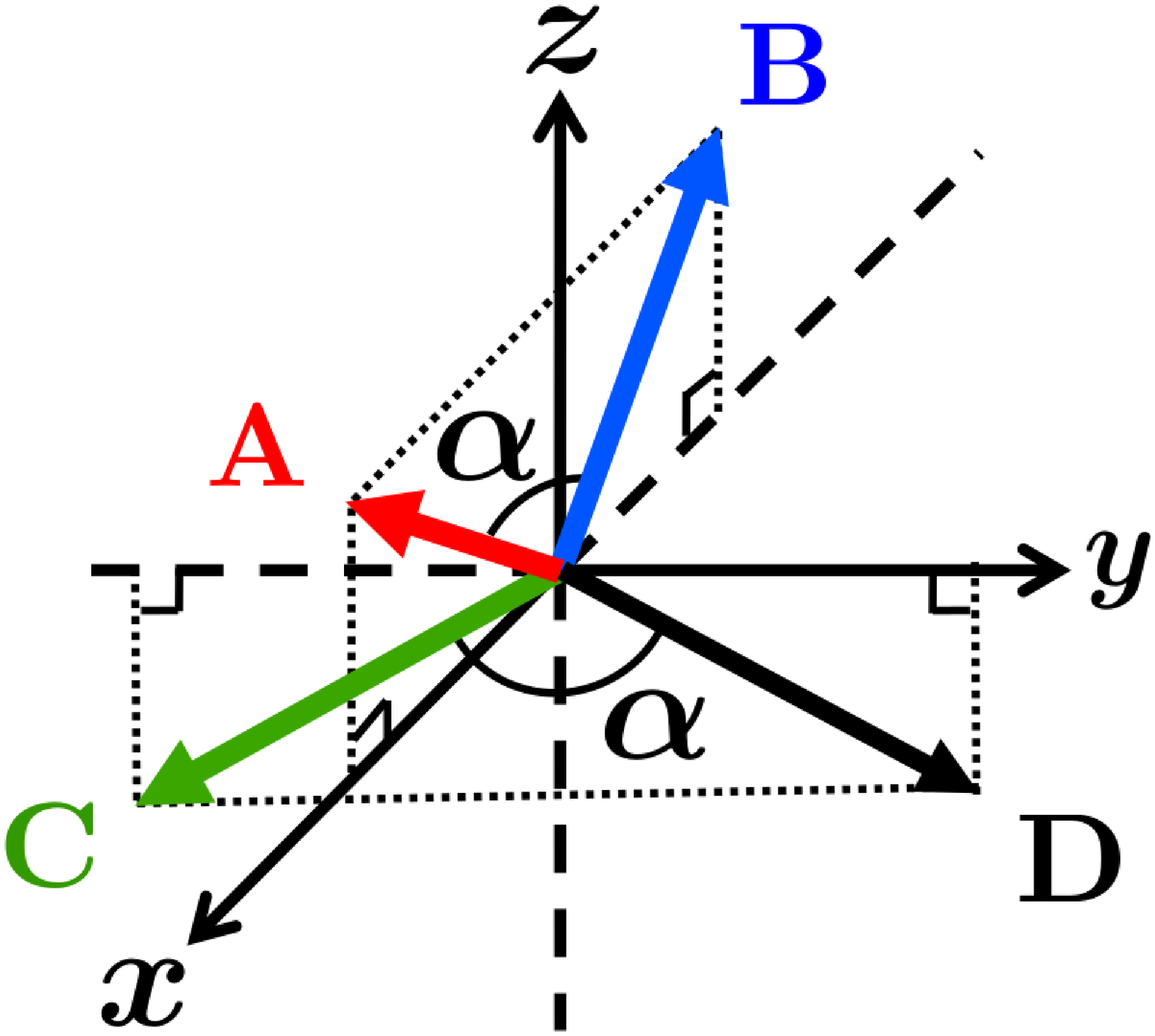}}}
	\caption{Example of (a) tetrahedral-structure state on a triangular lattice and (b) spin orientations of the state for three-dimensional coordinates.
	The four marks A, B, C, and D denote the four sublattices and $\mib{\delta}_{1}^{}$ and $\mib{\delta}_{2}^{}$ are the unit vectors.
	The solid, dotted, and broken lines in the upper panel denote the first-, second-, and third-nearest-neighbor interactions, respectively.
	One example is drawn for the second- and third-nearest-neighbor interactions.
	The angle $\alpha$ between the spin vectors satisfies $\cos{\alpha} = -1 / 3$.
	The spin vectors of sublattices A and B are in the $xz$-plane and those of sublattices C and D are in the $yz$-plane.}
	\label{tetrahedral}
\end{figure}

To investigate the MSE model on the basis of the linear spin-wave theory, first, we determine the region of the tetrahedral-structure phase of the classical ground state. 
The phase diagram of the classical ground state of the MSE model with up to the four-spin exchange interactions on the triangular lattice has already been estimated~\cite{Kubo1997} within the mean-field approximation assuming 144 sublattices.
The phase diagram is shown in Fig.~\ref{j_phase_diagram}.
There are the ferromagnetic phase for $J/K < -8.61$, the intermediate-phase region including various nearly degenerate phases with 12-, 18-, 24-, 72-, and 144-sublattice structures for $-8.61 \le J/K < -2.26$, the tetrahedral-structure phase for $-2.26 \le J/K < 8.22$, the six-sublattice-structure phase for $8.22 \le J/K < 10$, and the 120$^{\circ}$-structure phase for $J/K \ge 10$.
While the ferromagnetic state is stable for a large negative exchange integral of the two-spin exchange interaction, the 120$^{\circ}$-structure state is stable for a large positive exchange integral.
The 120$^{\circ}$-structure state has a finite vector chiral long-range order (LRO), the tetrahedral-structure state has a scalar chiral LRO, and the six-sublattice-structure state has both vector and staggered scalar chiral LRO~\cite{Yasuda2007}.
The spin structure of the tetrahedral-structure state consists of four sublattices, where the spin vectors on the four sublattices point to the four vertices of a tetrahedron if their bottoms are put at its center, as shown in Fig.~\ref{tetrahedral}.
The four spin vectors are at an angle $\alpha$ to each other, where  $\cos{\alpha} = -1 / 3$.
The phase diagram of the ground state of the MSE model with up to the six-spin interactions on a triangular lattice in a magnetic field has already been estimated within the mean-field approximation assuming 36 sublattices,~\cite{Yasuda2018} although the values of the five- and six-spin interactions have had limited investigation.

In the present work, we estimate the region of the tetrahedral-structure phase of the system within the mean-field approximation by the conjugate gradient (CG) method on $l \times l$ lattices with $l = 6$ and 12.
In the CG method, we prepare approximately $10^{2}~\mbox{--}~10^{5}$ initial states depending on the parameters.
We apply the tetrahedral structure and random states as the initial states of the CG method and find the smallest-energy state among the results of multiple calculations.
The values of the phase-transition points between the tetrahedral-structure phase and the others are estimated using the ground-state energy and scalar chirality
\begin{align}
\hat{\kappa}^{\rm s}
	= \sum_{\triangle} \mib{\sigma}_{i}^{} \cdot (\mib{\sigma}_{j}^{} \times \mib{\sigma}_{k}^{})~,
	\label{schiral}
\end{align}
where the summation of $\triangle$ is taken over all upward-pointing triangles on the triangular lattice.
In the tetrahedral-structure phase, the classical ground-state energy per site obeys
\begin{align}
\frac{E}{N}
	= - \Bigg( J + J_{\rm 2nd}^{} - 3J_{\rm 3rd}^{} + \frac{17}{3}K + 6L + \frac{59}{27} M \Bigg)~,
	\label{g-ene_j5j6}
\end{align}
and the value of the scalar chirality per site is $\langle \hat{\kappa}^{\rm s} \rangle / N = 4 / 3\sqrt{3} \simeq 0.77$, where $N$ is the number of sites.

\subsection{$J$-$K$-$L$-$M$ model with $J_{\rm 2nd}^{} = J_{\rm 3rd}^{} = 0$}

\begin{figure}[t]
	\centering
	\resizebox{0.45\textwidth}{!}{\includegraphics{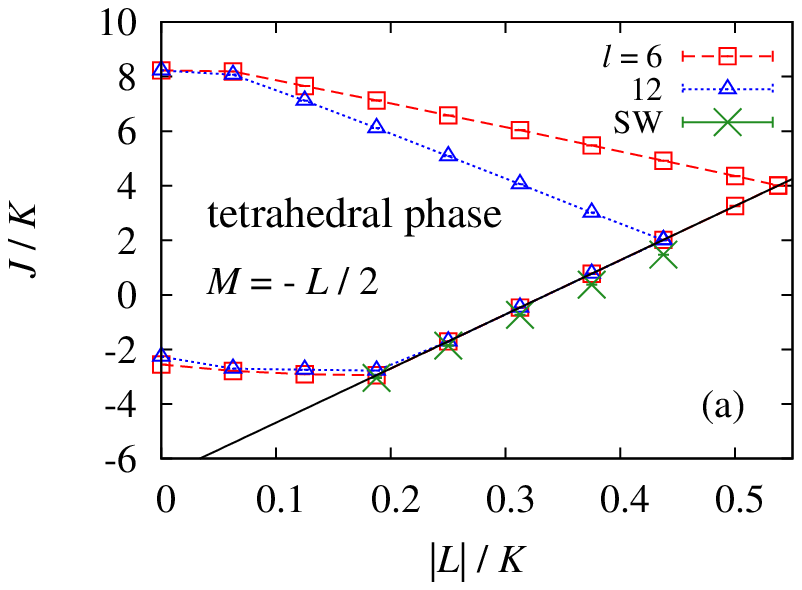}}\\
	\resizebox{0.45\textwidth}{!}{\includegraphics{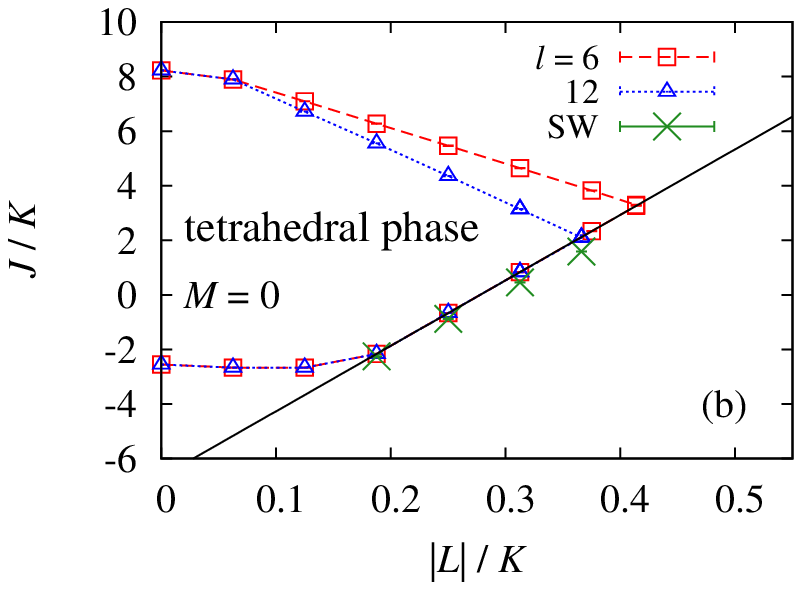}}
	\resizebox{0.45\textwidth}{!}{\includegraphics{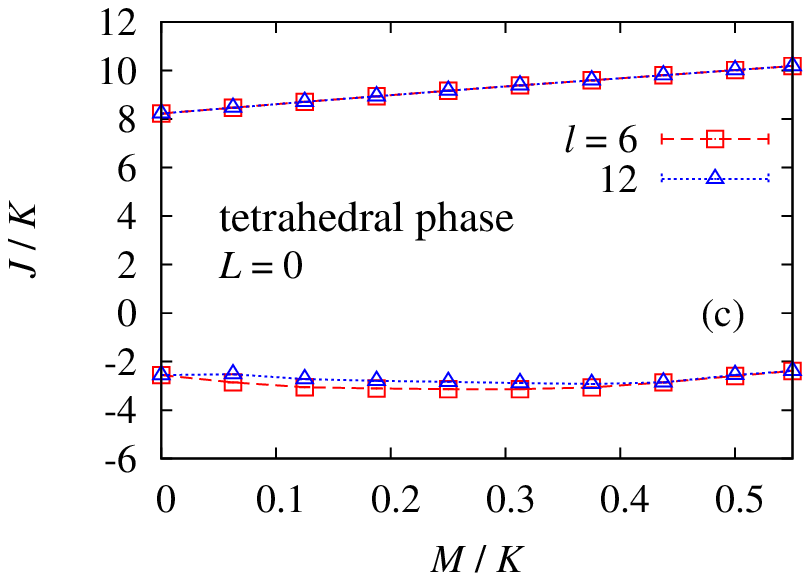}}
	\caption{Phase diagrams of the classical ground state for (a) $M = -L / 2$, (b) $M = 0$, and (c) $L = 0$ with $J_{\rm 2nd}^{} = J_{\rm 3rd}^{} = 0$.
	The squares and triangles are the phase-transition points calculated by the CG method for $l = 6$ and 12, respectively.
	The solid line is the boundary between the ferromagnetic and tetrahedral-structure phases estimated using Eqs.~(\ref{g-ene_j5j6}) and (\ref{ene_ferro}).
	The crosses labeled by SW are the phase-transition points estimated using the crossing points of the energy curves, which are obtained on the basis of the spin-wave theory for the ferromagnetic and tetrahedral-structure phases.
	The broken and dotted lines are guides to the eyes.}
	\label{phase_j5j6}
\end{figure}
We show the phase diagrams of the classical ground state for three cases, $M = -L / 2$ $(J_{5}^{} = J_{6}^{})$, $M = 0$, and $L = 0$ with $J_{\rm 2nd}^{} = J_{\rm 3rd}^{} = 0$, in Fig.~\ref{phase_j5j6}.
The squares and triangles in Fig.~\ref{phase_j5j6} are the phase-transition points calculated by the CG method for $l = 6$ and 12, respectively.
The region of the tetrahedral-structure phase is contracted for large $| L | / K$ and $M / K$, as shown in Fig.~\ref{phase_j5j6}(a). 
For $M = 0$, the region of the tetrahedral phase has the same shape as that of the $M = -L / 2$ system, but the maximum value of $| L | / K$ at which the system is in the tetrahedral phase becomes smaller than that of the $M = -L / 2$ system, as shown in Fig.~\ref{phase_j5j6}(b).
On the other hand, for $L = 0$, no significant change in the region of the tetrahedral phase with $M / K$ is observed in Fig.~\ref{phase_j5j6}(c).
These results show that the six-spin exchange interactions have little influence on the tetrahedral phase and the five-spin exchange interactions make the tetrahedral state unstable in this system.

For $| L | / K = M / K = 0$, it has been found that the phases with 12-, 18-, 72-, and 144-sublattice structures exist for $-8 \leq J / K \leq -2.26$ and that a six-sublattice-structure phase exists for $8.22 \leq J / K \leq 10$ in the $l = 12$ system~\cite{Kubo1997}.
The ferromagnetic-state energy $E_{\rm ferro}^{}$ obtained by the mean-field approximation is written as
\begin{align}
\frac{E_{\rm ferro}^{}}{N}
	= 3J + 21K + 90L + 31M~.
	\label{ene_ferro}
\end{align}
The solid lines in Fig.~\ref{phase_j5j6} are the boundaries between the ferromagnetic and tetrahedral-structure phases calculated using Eqs.~(\ref{g-ene_j5j6}) and (\ref{ene_ferro}).
The agreement of the triangles and the solid line shows that the ferromagnetic phase exists just under the tetrahedral phase in the range $0.25 \leq |L| / K \leq 0.4375$ and $0.1875 \leq |L| / K \leq 0.35$ for $M = -L/2$ and $M = 0$, respectively.
The crosses in Figs.~\ref{phase_j5j6}(a) and \ref{phase_j5j6}(b) are the phase-transition points estimated using the crossing points of the energy curves, which are obtained on the basis of the spin-wave theory for the ferromagnetic and tetrahedral-structure phases.
The details of the results obtained on the basis of the spin-wave theory are explained in Sect.~5.1.
In the adjacent phases of the tetrahedral-structure phase, except for those we explained above, the six- and twelve-sublattice-structure states and the states with various magnetizations are respectively realized in the upper and lower parts of the phase diagram shown in Fig.~\ref{phase_j5j6}.
As an exception, the umbrella state is realized for a large $M/K$ in the upper part of the phase diagram in Fig.~\ref{phase_j5j6}(c). 
The detailed analysis of the states around the tetrahedral-structure phase is beyond the scope of this work.

\subsection{$J$-$K$-$J_{\rm 2nd}^{}$-$J_{\rm 3rd}^{}$ model with $L = M = 0$}

\begin{figure}[t]
	\resizebox{0.45\textwidth}{!}{\includegraphics{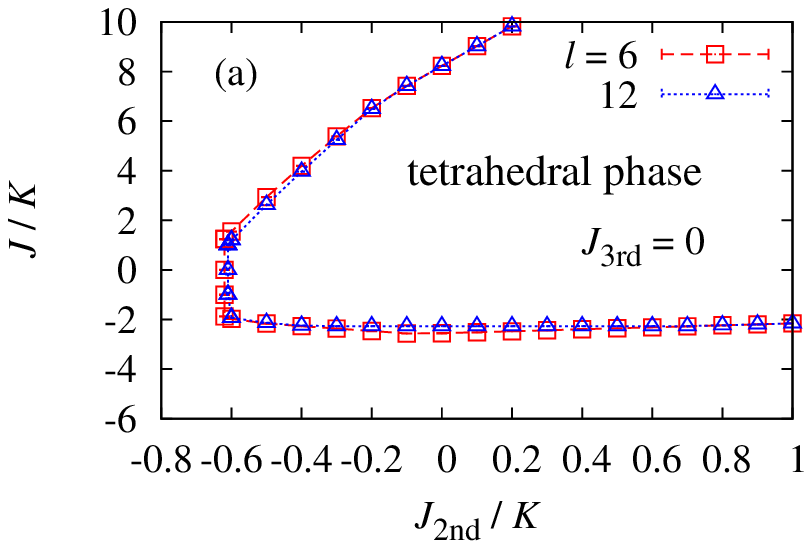}}
	\hspace{0.5cm}
	\resizebox{0.45\textwidth}{!}{\includegraphics{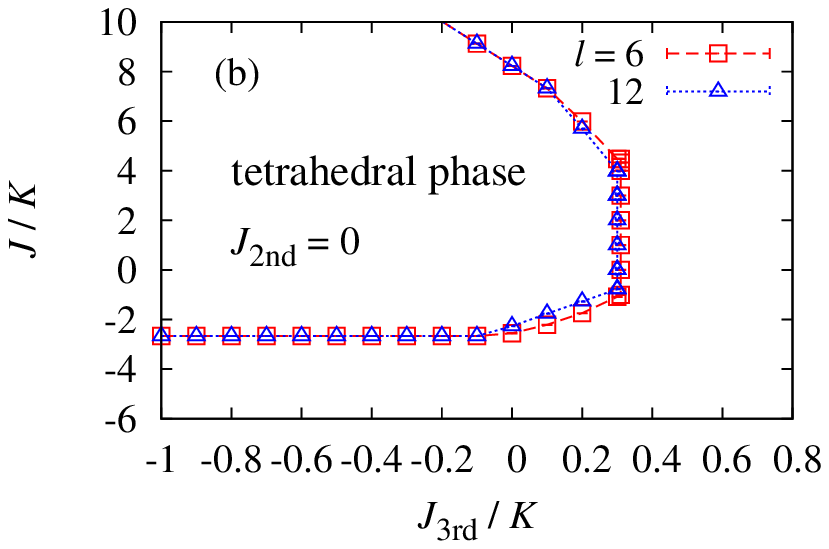}}
	\caption{Phase diagrams of the classical ground state for (a) $J_{\rm 3rd}^{} = 0$ and (b) $J_{\rm 2nd}^{} = 0$ with $L = M = 0$.
	The squares and triangles are the phase-transition points calculated by the CG method for $l = 6$ and 12, respectively.
	The lines are guides to the eyes.}
	\label{phase}
\end{figure}

In this work, we also study the effects of the second- and third-nearest-neighbor interactions in addition to the five- and six-spin exchange interactions.
We show the phase diagrams of the classical ground state for two cases, $J_{\rm 3rd}^{} = 0$ and $J_{\rm 2nd}^{} = 0$ with $L = M = 0$, in Fig.~\ref{phase}.
For $J_{\rm 3rd}^{} = 0$, the region of the tetrahedral-structure phase is expanded for antiferromagnetic $J_{\rm 2nd}^{} / K> 0$ and is contracted for ferromagnetic $J_{\rm 2nd}^{} / K < 0$, as shown in Fig.~\ref{phase}(a).
Namely, the antiferromagnetic second-nearest-neighbor interactions stabilize the tetrahedral-structure state within the mean-field approximation.
The result is consistent with an expectation from the situation that the two sites connected by $J_{\rm 2nd}^{}$ always belong to different sublattices.
While the lower line of the phase boundary is robust against the change in $J_{\rm 2nd}^{} / K$, the upper line is sensitive to the change in $J_{\rm 2nd}^{} / K$.
The states below the lower line of the phase diagram have various structures consisting of many sublattices~\cite{Kubo1997}.
While the phase above the upper line has a six-sublattice structure at $J_{\rm 2nd}^{} = 0$~\cite{Kubo1997}, we find that, for $J_{\rm 2nd}^{} / K \leq -0.2$, the spin configurations calculated do not have a six-sublattice structure but a twelve-sublattice structure.
The upper and lower lines connect at $J_{\rm 2nd}^{} / K \simeq -0.6$.
Because a detailed further analysis is beyond the scope of this work, we do not precisely estimate the boundary and the states around the tetrahedral-structure phase.

For $J_{\rm 2nd}^{} = 0$, the region of the tetrahedral-structure phase is contracted for antiferromagnetic $J_{\rm 3rd}^{} / K > 0$ and is expanded for ferromagnetic $J_{\rm 3rd}^{} / K < 0$ as shown in Fig.~\ref{phase}(b).
Namely, the ferromagnetic third-nearest-neighbor interactions stabilize the tetrahedral-structure state within the mean-field approximation.
The result is consistent with an expectation from the situation that the two sites connected by $J_{\rm 3rd}^{}$ always belong to the same sublattices.
Similar to the case of $J_{\rm 3rd}^{} = 0$, the lower line of the phase transition is robust against the change in $J_{\rm 3rd}^{} / K$, and the upper line is sensitive to the change in $J_{\rm 3rd}^{} / K$.
The states below the lower line of the phase diagram have various structures consisting of many sublattices, and the phase above the upper line is a six-sublattice-structure phase for $J_{\rm 3rd}^{} / K < 0.2$ and a twelve-sublattice-structure phase for $J_{\rm 3rd}^{} / K \geq 0.2$.
The upper and lower lines connect at $J_{\rm 3rd}^{} / K \simeq 0.3$.
Because a detailed further analysis is beyond the scope of this work, we do not precisely estimate the boundary and the states around the tetrahedral-structure phase.


\section{Spin-Wave Theory}

We examine the stability of the tetrahedral-structure state against quantum fluctuations using the linear spin-wave theory based on the classical ground state.
The tetrahedral-structure state has a four-sublattice structure with four spins, as shown in Fig.~\ref{tetrahedral}(b).
Using the Holstein--Primakoff~\cite{Holstein1940} and Fourier transformations, we obtain the Hamiltonian
\begin{align}
\cal{H}
	&= -\Bigg( J + J_{\rm 2nd}^{} - 3J_{\rm 3rd}^{} + \frac{17}{3}K + 6L + \frac{59}{27}M \Bigg) N
	\notag \\
	&~~~~
	+ \sum_{\mib{k}} \Big[
	A(\mib{k}) \Big( a_{\mib{k}}^{\dag}a_{\mib{k}}^{} + b_{\mib{k}}^{\dag}b_{\mib{k}}^{} + c_{\mib{k}}^{\dag}c_{\mib{k}}^{} + d_{\mib{k}}^{\dag}d_{\mib{k}}^{} \Big)
	\notag \\
	&	~~~~
	+ B_{1}^{}(\mib{k}) \Big( a_{\mib{k}}^{\dag}b_{\mib{k}}^{} + c_{\mib{k}}^{\dag}d_{\mib{k}}^{} + \mathrm{h.c.} \Big)
	\notag \\
	&~~~~
	+ B_{2}^{}(\mib{k}) \Big( a_{\mib{k}}^{\dag}c_{\mib{k}}^{} + b_{\mib{k}}^{\dag}d_{\mib{k}}^{} + \mathrm{h.c.} \Big)
	\notag \\
	&~~~~
	+ B_{3}^{}(\mib{k}) \Big( a_{\mib{k}}^{\dag}d_{\mib{k}}^{} + b_{\mib{k}}^{\dag}c_{\mib{k}}^{} + \mathrm{h.c.} \Big)
	\notag \\
	&~~~~
	+ C_{1}^{}(\mib{k}) \Big( a_{-\mib{k}}^{}b_{\mib{k}}^{} + c_{-\mib{k}}^{}d_{\mib{k}}^{} + \mathrm{h.c.} \Big)
	\notag \\
	&~~~~
	+ C_{2}^{}(\mib{k}) \Big \{ \phi^{\ast} \Big( a_{-\mib{k}}^{}c_{\mib{k}}^{} + b_{-\mib{k}}^{}d_{\mib{k}}^{} \Big) + \mathrm{h.c.} \Big \}
	\notag \\
	&~~~~
	+ C_{3}^{}(\mib{k}) \Big \{ \phi \Big( a_{-\mib{k}}^{}d_{\mib{k}}^{} + b_{-\mib{k}}^{}c_{\mib{k}}^{} \Big) + \mathrm{h.c.} \Big \} 
	\notag \\
	&~~~~
	+ D (\mib{k}) \Big( a_{-\mib{k}}^{}a_{\mib{k}}^{} + b_{-\mib{k}}^{}b_{\mib{k}}^{} + c_{-\mib{k}}^{}c_{\mib{k}}^{} + d_{-\mib{k}}^{}d_{\mib{k}}^{} \Big)
	+ {\rm h.c.} \Big \}	
	\Big]~,
	\label{ham_Fourier_5-6-spin}
\end{align}
where
\begin{align}
A(\mib{k})
	&= \frac{4}{9} (9J + 9J_{\rm 2nd}^{} - 27J_{\rm 3rd}^{} + 48K + 32M)
	\notag \\
	&+ \frac{4}{27} (27J_{\rm 3rd}^{} + 24L + 16M) \{ \cos 2k_{2}^{} 
	\notag \\
	&+ \cos 2(k_{1}^{} - k_{2}^{}) + \cos 2k_{1}^{} \} ~,
	\notag \\
B_{1}(\mib{k})
	&= \frac{4}{27} (-9J - 12K - 24L + 32M) \cos k_{2}^{}
	\notag \\
	&+ \frac{4}{27} (-9J_{\rm 2nd}^{} + 12K - 16M) \cos (2k_{1}^{} - k_{2}^{})~,
	\notag \\
B_{2}(\mib{k})
	&= \frac{4}{27} (-9J - 12K - 24L + 32M) \cos (k_{1}^{} - k_{2}^{})
	\notag \\
	&	+ \frac{4}{27} (-9J_{\rm 2nd}^{} + 12K - 16M) \cos (k_{1}^{} + k_{2}^{})~,
	\notag \\
B_{3}(\mib{k})
	&= \frac{4}{27} (-9J - 12K - 24L + 32M) \cos k_{1}^{}
	\notag \\
	&	+ \frac{4}{27} (-9J_{\rm 2nd}^{} + 12K - 16M) \cos (-k_{1}^{} + 2k_{2}^{})~,
	\notag \\
C_{1}(\mib{k})
	&= \frac{8}{27} (9J + 30K + 24L + 16M) \cos k_{2}^{}
	\notag \\
	&	+ \frac{8}{27} (9J_{\rm 2nd}^{} + 6K + 16M) \cos (2k_{1}^{} - k_{2}^{})~,
	\notag \\
C_{2}(\mib{k})
	&= \frac{8}{27} (9J + 30K + 24L + 16M) \cos (k_{1}^{} - k_{2}^{})
	\notag \\
	&	+ \frac{8}{27} (9J_{\rm 2nd}^{} + 6K + 16M) \cos (k_{1}^{} + k_{2}^{})~,
	\notag \\
C_{3}(\mib{k})
	&= \frac{8}{27} (9J + 30K + 24L + 16M) \cos k_{1}^{}
	\notag \\
	&	+ \frac{8}{27} (9J_{\rm 2nd}^{} + 6K + 16M) \cos (-k_{1}^{} + 2k_{2}^{})~,
	\notag \\
D(\mib{k})
	&= -\frac{64}{27} (3L - M) \{ \cos 2k_{2} + \phi^{\ast} \cos 2(k_{1}^{} - k_{2}^{}) 
	\notag \\
	&+ \phi \cos 2k_{1}^{} \}~,
	\notag \\
\phi
	&= \exp(-2\pi i / 3)~.
	\label{coef_A_B_C_D}
\end{align}
Here, $\mib{k}=(k_1, k_2)$ are the wavevectors with $k_i= \mib{k} \cdot \mib{\delta}_i$, and $\mib{\delta}_1$ and $\mib{\delta}_2$ are the unit vectors chosen as shown in Fig.~\ref{tetrahedral}.
The operators $a_{\mib{k}}^{\dag}$, $b_{\mib{k}}^{\dag}$, $c_{\mib{k}}^{\dag}$, and $d_{\mib{k}}^{\dag}$ are, respectively, the creation operators for bosons on the four sublattices, and $a_{\mib{k}}^{}$, $b_{\mib{k}}^{}$, $c_{\mib{k}}^{}$, and $d_{\mib{k}}^{}$ are the annihilation operators of the bosons.

We can transform the Hamiltonian in Eq.~(\ref{ham_Fourier_5-6-spin}) with a unitary transformation and obtain
\begin{align}
\mathcal{H}
	&= -\Bigg( J + J_{\rm 2nd}^{} - 3J_{\rm 3rd}^{} + \frac{17}{3}K + 6L + \frac{59}{27}M \Bigg)N
	\notag \\
	&~~~~
	+ \sum_{\mu = 1}^{4} {\sum_{\mib{k}}}' \Big \{
	X_{\mu}^{}(\mib{k}) \Big( {\tilde a}_{\mu,\mib{k}}^{\dag}{\tilde a}_{\mu,\mib{k}}^{} + {\tilde a}_{\mu,-\mib{k}}^{\dag}{\tilde a}_{\mu,-\mib{k}}^{} \Big)
	\notag \\
	&~~~~
	+ Y_{\mu}^{}(\mib{k}){\tilde a}_{\mu,-\mib{k}}^{}{\tilde a}_{\mu,\mib{k}}^{}
	+ Y_{\mu}^{\ast}(\mib{k}){\tilde a}_{\mu,-\mib{k}}^{\dag}{\tilde a}_{\mu,\mib{k}}^{\dag} \Big \}~,
	\label{ham_unitary_5-6-spin}
\end{align}
where
\begin{align}
X_{1}^{}(\mib{k})
	&= A(\mib{k}) + B_{1}^{}(\mib{k}) + B_{2}^{}(\mib{k}) + B_{3}^{}(\mib{k})~,
	\notag \\
X_{2}^{}(\mib{k})
	&= A(\mib{k}) + B_{1}^{}(\mib{k}) - B_{2}^{}(\mib{k}) - B_{3}^{}(\mib{k})~,
	\notag \\
X_{3}^{}(\mib{k})
	&= A(\mib{k}) - B_{1}^{}(\mib{k}) + B_{2}^{}(\mib{k}) - B_{3}^{}(\mib{k})~,
	\notag \\
X_{4}^{}(\mib{k})
	&= A(\mib{k}) - B_{1}^{}(\mib{k}) - B_{2}^{}(\mib{k}) + B_{3}^{}(\mib{k})~,
	\notag \\
Y_{1}^{}(\mib{k})
	&= D(\mib{k}) + C_{1}^{}(\mib{k}) + \phi^{\ast} C_{2}^{}(\mib{k}) + \phi C_{3}^{}(\mib{k})~,
	\notag \\
Y_{2}^{}(\mib{k})
	&= D(\mib{k}) + C_{1}^{}(\mib{k}) - \phi^{\ast} C_{2}^{}(\mib{k}) - \phi C_{3}^{}(\mib{k})~,
	\notag \\
Y_{3}^{}(\mib{k})
	&= D(\mib{k}) - C_{1}^{}(\mib{k}) + \phi^{\ast} C_{2}^{}(\mib{k}) - \phi C_{3}^{}(\mib{k})~,
	\notag \\
Y_{4}^{}(\mib{k})
	&= D(\mib{k}) - C_{1}^{}(\mib{k}) - \phi^{\ast} C_{2}^{}(\mib{k}) + \phi C_{3}^{}(\mib{k})~,
	\label{coeff_X_Y}
\end{align}
and the summation $\sum_{\mib{k}}'$ is taken over the wave numbers in the half-region of the Brillouin zone.
The boson operators ${\tilde a}_{\mu,\mib{k}}^{\dag}$ are defined by
\begin{align}
\left [
	\begin{array}{cccc}
	{\tilde a}_{1,\mib{k}}^{\dag} \\
	{\tilde a}_{2,\mib{k}}^{\dag} \\
	{\tilde a}_{3,\mib{k}}^{\dag} \\
	{\tilde a}_{4,\mib{k}}^{\dag}
	\end{array}
\right ]
	&= \frac{1}{2} \left [
		\begin{array}{cccc}
		1 & 1 & 1 & 1 \\
		1 & 1 & -1 & -1 \\
		1 & -1 & 1 & -1 \\
		1 & -1 & -1 & 1 
		\end{array}
	\right ]
	\left [
		\begin{array}{cccc}
		a_{\mib{k}}^{\dag} \\
		b_{\mib{k}}^{\dag} \\
		c_{\mib{k}}^{\dag} \\
		d_{\mib{k}}^{\dag}
		\end{array}
	\right ] \ .
\end{align}
Finally, transforming the Hamiltonian in Eq.~(\ref{ham_unitary_5-6-spin}) with a Bogoliubov transformation, we obtain
\begin{align}
\cal{H}
	&= -\Bigg( J + J_{\rm 2nd}^{} - 3J_{\rm 3rd}^{} + \frac{17}{3}K + 6L + \frac{59}{27}M \Bigg) N
	\notag \\
	&~~~~
	+ \frac{1}{2} \sum_{\mu = 1}^{4} \sum_{\mib{k}} \{ \omega_{\mu}(\mib{k})^{} - X_{\mu}(\mib{k})^{} \}
	\notag \\
	&~~~~
	+ \sum_{\mu = 1}^{4} \sum_{\mib{k}} \omega_{\mu}(\mib{k})^{} \alpha_{\mu,\mib{k}}^{\dag}\alpha_{\mu,\mib{k}}^{}~,
	\label{ham_Bogoliubov_5-6-spin}
\end{align}
where the spin-wave frequency is written as
\begin{align}
\omega_{\mu}(\mib{k})^{}
	= \sqrt{ \{ X_{\mu}(\mib{k})^{} \} ^{2} - | Y_{\mu}(\mib{k})^{} |^{2}}~,
	\label{omega-tilde}
\end{align}
and $\alpha_{\mu,\mib{k}}^{\dag}$ and $\alpha_{\mu,\mib{k}}^{}$ are the creation and annihilation operators of spin waves with $\mu$-modes, respectively.

\begin{figure}[t]
	\centering
	\resizebox{0.45\textwidth}{!}{\includegraphics{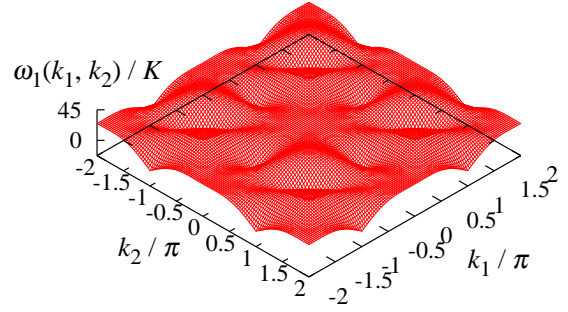}}
	\caption{Wave-number dependences of $\omega_{1}^{}(k_{1}^{},k_{2}^{}) / K$ for $J / K = 2$, $L / K = -0.25$, and $M / K = 0.125$ with $J_{\rm 2nd}^{} = J_{\rm 3rd}^{} = 0$.}
	\label{omega_3d_j5j6}
\end{figure}
We show the wave-number dependences of $\omega_{1}^{}(k_{1}^{},k_{2}^{}) / K$ for $J / K = 2$, $L / K = -0.25$, $M / K = 0.125$, and $J_{\rm 2nd}^{} = J_{\rm 3rd}^{} = 0$ in Fig.~\ref{omega_3d_j5j6}.
Note that the system with $L / K = -0.25$ and $M / K = 0.125$ corresponds to that with $J_{5}^{} / J_{4}^{} = J_{6}^{} / J_{4}^{} = 0.5$.
The dispersion relation of the spin-wave frequency $\omega_1(\mib{k})$ has a sixfold rotational symmetry.
The spin-wave frequencies of the four branches are expressed with a single analytical form, i.e., we have the relations
\begin{align}
\omega_{1}^{}(k_{1}^{}, k_{2}^{})
	&= \omega_{2}^{}(k_{1}^{} - \pi, k_{2}^{})
	= \omega_{3}^{}(k_{1}^{} - \pi, k_{2}^{} - \pi)
	\notag \\
	&
	= \omega_{4}^{}(k_{1}^{}, k_{2}^{} - \pi)~.
	\label{omega_1-4}
\end{align}
The wave numbers at which the values of the energy frequency $\omega_{1}^{}(k_{1}^{}, k_{2}^{})$ equal zero are $(\pm \pi, 0)$, $(0,\pm \pi)$, and $(\pm \pi, \pm \pi)$, and the frequency $\omega_{1}^{}(0, 0)$ at $\mib{k} = \mib{0}$ is finite, i.e., gapful.
The frequency is written as
\begin{align}
\omega_{1}^{}(k_{1}^{},k_{2}^{})
	&\simeq \frac{64}{9} (3K + 4M) 
	\notag \\
	&
	+ \frac{4}{9} (3J + 9J_{\rm 2nd}^{} - 36J_{\rm 3rd}^{} - 8K - 24L - 16M)
	\notag \\
	&
 	\times (k_{1}^{2} - k_{1}^{}k_{2}^{} + k_{2}^{2})
	\label{omega_1}
\end{align}
for small $k$ ($k=|\mib{k}|$).
The value of $\omega_{1}^{}(k_{1}^{},k_{2}^{})$ at $\mib{k} = \mib{0}$ depends on $M$ and is independent of $L$.
On the other hand, $\omega_{\mu}^{}(k_{1}^{}, k_{2}^{})~(\mu = 2, 3, 4)$ are gapless and their energy frequencies for small $k$ are proportional to $k$.
The energy frequency for mode 2 is written as
\begin{align}
\omega_{2}^{}(k_{1}^{}, k_{2}^{})
	&\simeq \frac{8}{27} \{(9J + 9J_{\rm 2nd}^{} + 36K + 24L + 32M)
	\notag \\
	&
	\times (u_{1}^{} k_{1}^{2} - u_{1}^{} k_{1}^{} k_{2}^{} + u_{2}^{} k_{2}^{2}) \}^{1/2}
	\label{omega_2}
\end{align}
with
$u_{1}^{} = 12 ( 9J_{\rm 2nd}^{} - 18J_{\rm 3rd}^{} + 6K + 8M )$
and
$u_{2}^{} = 9 ( 3J + 3J_{\rm 2nd}^{} - 24J_{\rm 3rd}^{} + 20K - 24L )$
for small $k$.
From the positional relationship of the zero points for mode 2, 3, and 4 frequencies, we obtain the relations
\begin{align}
&\omega_{4}^{}(k_{1}^{},k_{2}^{})
	= \omega_{2}^{}(k_{1}^{} - k_{2}^{},k_{1}^{})~,
	\notag \\
	&
\omega_{3}^{}(k_{1}^{},k_{2}^{})
	= \omega_{4}^{}(k_{1}^{} - k_{2}^{},k_{1}^{})~.
	\label{omega_2-3-4}
\end{align}

\begin{figure}[t]
	\centering
	\resizebox{0.45\textwidth}{!}{\includegraphics{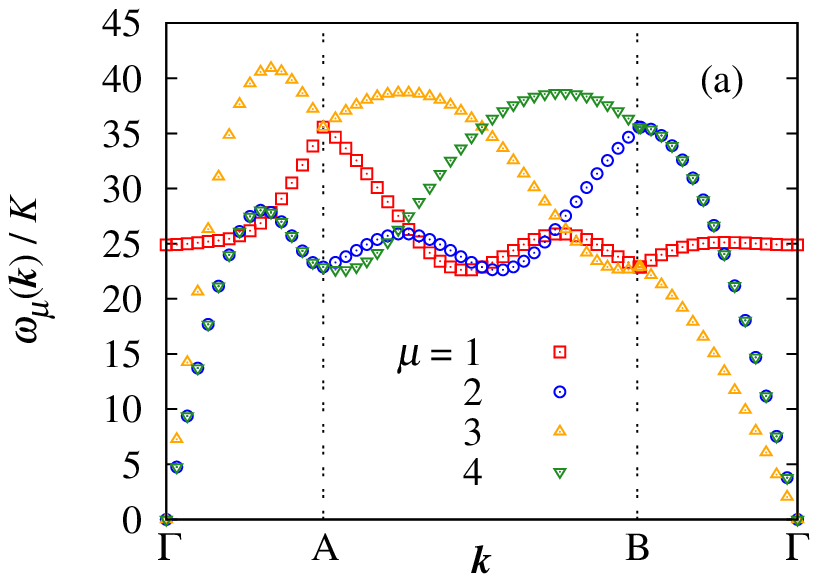}}
	\resizebox{0.45\textwidth}{!}{\includegraphics{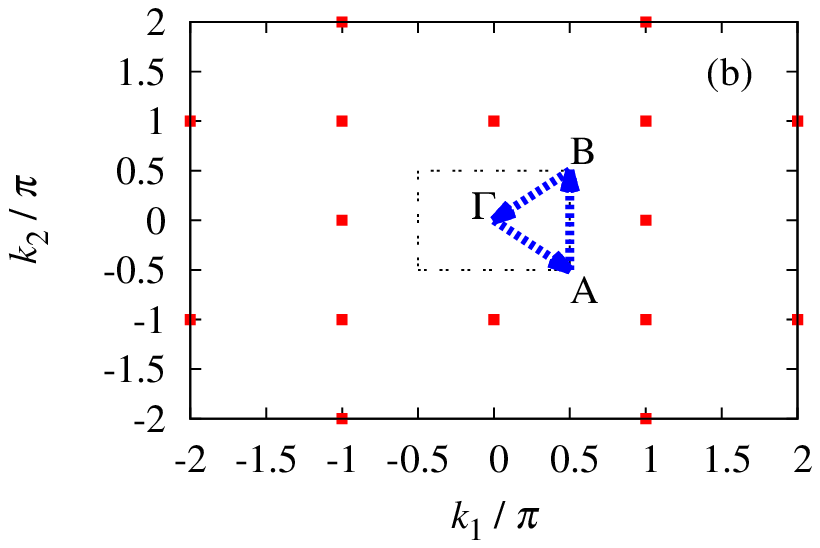}}
	\caption{(a) Wave-number dependences of $\omega_{\mu}^{}(\mib{k}) / K$ on the $\Gamma$-A-B-$\Gamma$ lines for $J / K = 2$, $L / K = -0.25$, and $M / K = 0.125$ with $J_{\rm 2nd}^{} = J_{\rm 3rd}^{} = 0$, and (b) $\Gamma$-A-B-$\Gamma$ lines in the first Brillouin zone drawn with the double-dotted-line rectangle.
	The translational vectors in real space are taken to be $\mib{\delta}_{1}^{} = (1, 0)$ and $\mib{\delta}_{2}^{} = (0, 1)$.
	The bold squares in the lower panel denote the zero points of the energy frequency $\omega_{1}^{}(\mib{k})$.}
	\label{omega_2d_j5j6}
\end{figure}

We show the wave-number dependences of $\omega_{\mu}^{}(\mib{k}) / K$ on the $\Gamma$-A-B-$\Gamma$ lines for $J / K = 2$, $L / K = -0.25$, and $M / K = 0.125$ with $J_{\rm 2nd}^{} = J_{\rm 3rd}^{} = 0$ in Fig.~\ref{omega_2d_j5j6}(a).
Here, the $\Gamma$-A-B-$\Gamma$ lines with $\Gamma=(0, 0)$, ${\rm A}=\pi/2(1, -1)$, and ${\rm B}=\pi/2(1, 1)$ exist in the first Brillouin zone, as shown in Fig.~\ref{omega_2d_j5j6}(b).
The translational vectors in real space are taken to be $\mib{\delta}_{1}^{} = (1, 0)$ and $\mib{\delta}_{2}^{} = (0, 1)$, as shown in Fig.~\ref{tetrahedral}(a).
The bold squares in Fig.~\ref{omega_2d_j5j6}(b) are the zero points of the energy frequency $\omega_{1}^{}(\mib{k})$.
The values of $\omega_{\mu}^{}(\mib{k}) / K$ increase with $J/K$ for fixed $L / K$ and $M / K$.
Furthermore, as far as we can see in the spectra, there is no significant change when the values of $L / K$ and $M / K$ are changed for a fixed $J / K$.
If the spectra of the spin waves show softening, a phase transition occurs.
We find that none of the spectra soften for any $\mib{k}$ in the whole region of the tetrahedral-structure phase of the classical system.
Softening is observed outside the tetrahedral-structure phase.

\begin{figure}[t]
	\centering
	\resizebox{0.45\textwidth}{!}{\includegraphics{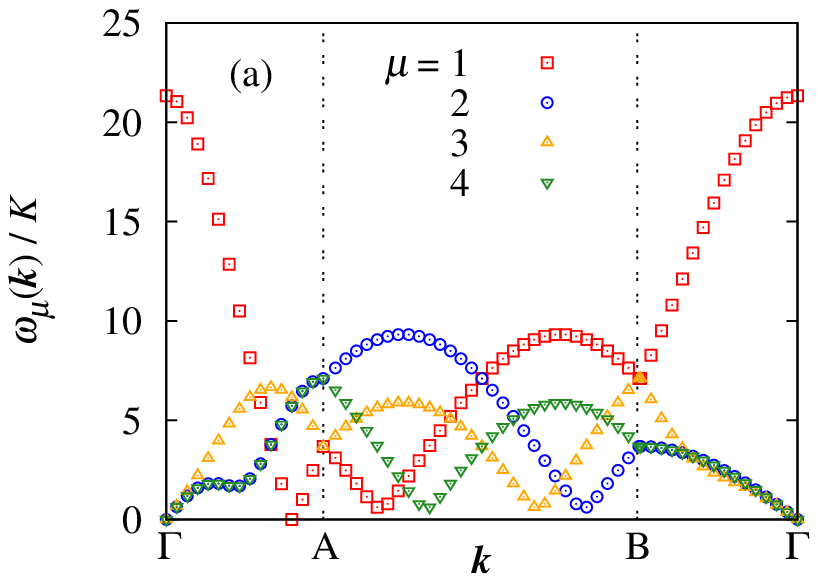}}
	\resizebox{0.45\textwidth}{!}{\includegraphics{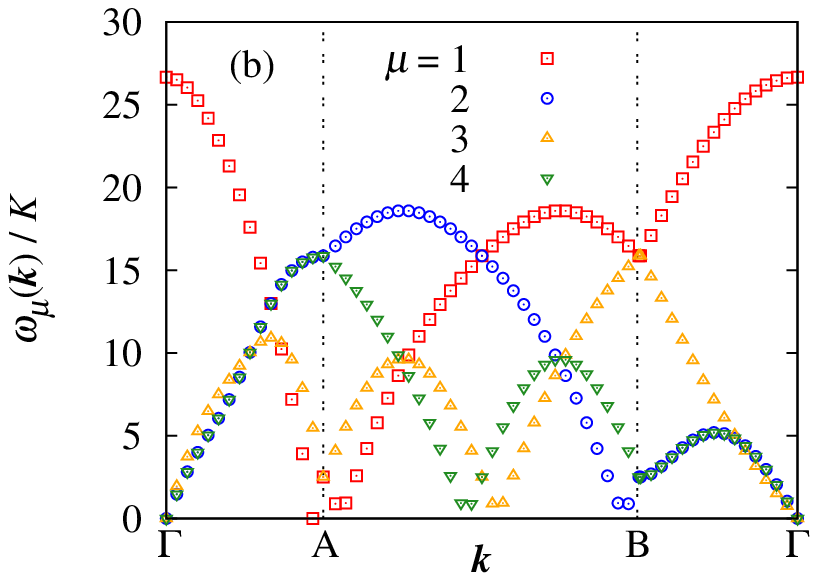}}
	\caption{Wave-number dependences of $\omega_{\mu}^{}(\mib{k}) / K$ on the $\Gamma$-A-B-$\Gamma$ lines for (a) $J / K = -3.78$ with $L = M = 0$ and (b) $J / K = -2.96$ with $L / K = -0.375$ and $M / K = 0.1875$ in the $J$-$K$-$L$-$M$ system.}
	\label{soft}
\end{figure}
We investigate the softening for three cases, $M = -L/2$, $M = 0$, and $L = 0$ with $J_{\rm 2nd}^{} = J_{\rm 3rd}^{} = 0$.
As an example, we show the wave-number dependences of $\omega_{\mu}^{}(\mib{k}) / K$ for $J / K = -3.78$ with $L = M = 0$ and for $J / K = -2.96$ with $L / K = -0.375$ and $M / K = 0.1875$ in Figs.~\ref{soft}(a) and \ref{soft}(b), respectively.
The frequency $\omega_1(\mib{k})$ decreases with increasing ferromagnetic coupling $| J | / K$ and becomes zero near $\mib{k} \simeq \pi/2(0.8, -0.8)$, and  $\pi/2(0.9, -0.9)$ for $J / K \simeq -3.78$ and $-2.96$ in Figs.~\ref{soft}(a) and \ref{soft}(b), respectively.
The results for other parameter sets show the appearance of softening at various $\mib{k}$ for various $J / K$ close to $J / K = -4$.

We also investigate the wave-number dependences of the spectra for two cases, $J_{\rm 3rd}^{} = 0$ and $J_{\rm 2nd}^{} = 0$ with $L = M = 0$.
The energy frequencies $\omega_{\mu}^{}(\mib{k})$ increase with $J/K$ for fixed $J_{\rm 2nd}^{} / K$ and $J_{\rm 3rd}^{} / K$.
As far as we can see in the spectra, there is no significant change when the values of $J_{\rm 2nd}^{} / K$ and $J_{\rm 3rd}^{} / K$ are changed for a fixed $J / K$.
We find that none of the spectra soften for any $\mib{k}$ in the whole region of the tetrahedral-structure phase of the classical system.
Softening is observed outside the tetrahedral-structure phase.
Similar to the $J$-$K$-$L$-$M$ system, softening appears at various $\mib{k}$ for various $J / K$ close to $J / K = -4$.
Because these instabilities observed in the $J$-$K$-$L$-$M$ and $J$-$K$-$J_{\rm 2nd}^{}$-$J_{\rm 3rd}^{}$ systems occur outside the mean-field phase, they are beyond the scope of this study.


\section{Physical Quantities}

In this section, we describe physical quantities in the $J$-$K$-$L$-$M$ and $J$-$K$-$J_{\rm 2nd}^{}$-$J_{\rm 3rd}^{}$ models.
Note that the results outside the tetrahedral-structure phase are also plotted in figures where we show the $J / K$ dependences of the physical quantities.

\subsection{$J$-$K$-$L$-$M$ model}

First, we set the parameters $J_{\rm 2nd}^{} = J_{\rm 3rd}^{} = 0$ and describe physical quantities in the $J$-$K$-$L$-$M$ model.

\subsubsection{Ground-state energy}

\begin{figure}[t]
	\centering
	\resizebox{0.45\textwidth}{!}{\includegraphics{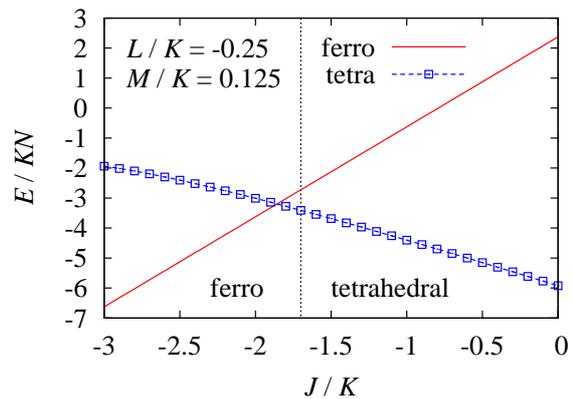}}
	\caption{$J / K$ dependences of the ground-state energies of the ferromagnetic and tetrahedral-structure states for $L / K = -0.25$ and $M / K = 0.125$.
	The squares are the results for the tetrahedral-structure state obtained on the basis of the spin-wave theory, and the solid line is the analytical result for the ferromagnetic state.
	The vertical dotted line at $J / K = -1.7$ is the classical phase boundary between the ferromagnetic and tetrahedral-structure phases.
	The broken line is a guide to the eyes.}
	\label{ene_ferro-tetra}
\end{figure}

To compare the ground-state energies of the ferromagnetic state with those of the tetrahedral-structure state obtained on the basis of the spin-wave theory, we show the $J / K$ dependences of the ground-state energies for $L / K = -0.25$ and $M / K = 0.125$ in Fig.~\ref{ene_ferro-tetra}.
The squares are the results of the tetrahedral-structure state obtained on the basis of the spin-wave theory, and the solid line is the analytical result of the ferromagnetic state [Eq.~(\ref{ene_ferro})].
The phase-transition point evaluated by the crossing point in Fig.~\ref{ene_ferro-tetra} is at $J / K \simeq -1.86$.
The vertical dotted line at $J / K = -1.7$ is the classical phase boundary between the ferromagnetic and tetrahedral-structure phases.
The phase-transition point in $J / K$ obtained on the basis of the spin-wave theory is smaller than that obtained by the mean-field approximation.
The result shows that the tetrahedral-structure state is stabilized by the quantum fluctuations.
The phase-transition points obtained with the crossing points of the energies for $0.1875 \leq | L | / K \leq 0.4375$ are denoted by the crosses in Figs.~\ref{phase_j5j6}(a) and \ref{phase_j5j6}(b).
The phase-transition line between the tetrahedral-structure and ferromagnetic phases slightly shifts to the side of larger ferromagnetic coupling $| J | / K$.
The change is larger for a large $|L| / K$.
In order to investigate the phase-transition boundaries, except for the ferromagnetic phase, we need to perform the spin-wave analysis for various phases, e.g., the twelve-sublattice structure.
In the spin-wave theory for the MSE model with up to the six-spin interactions, because an enormous number of calculations were required even for the four-sublattice structure, some ideas are needed for the six- and twelve-sublattice structures.
This is beyond the scope of this work.

\subsubsection{Quantum correction of the ground-state energy}

\begin{figure}[t]
	\centering
	\resizebox{0.45\textwidth}{!}{\includegraphics{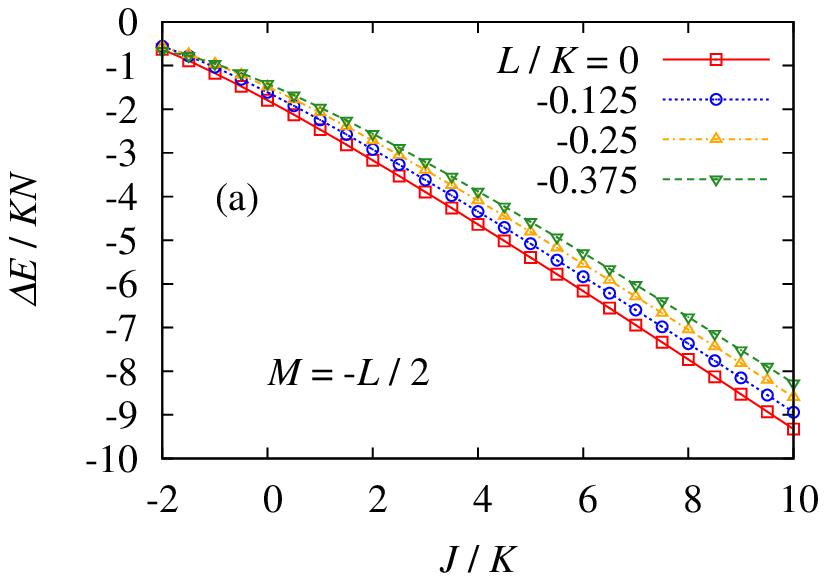}}\\
	\resizebox{0.45\textwidth}{!}{\includegraphics{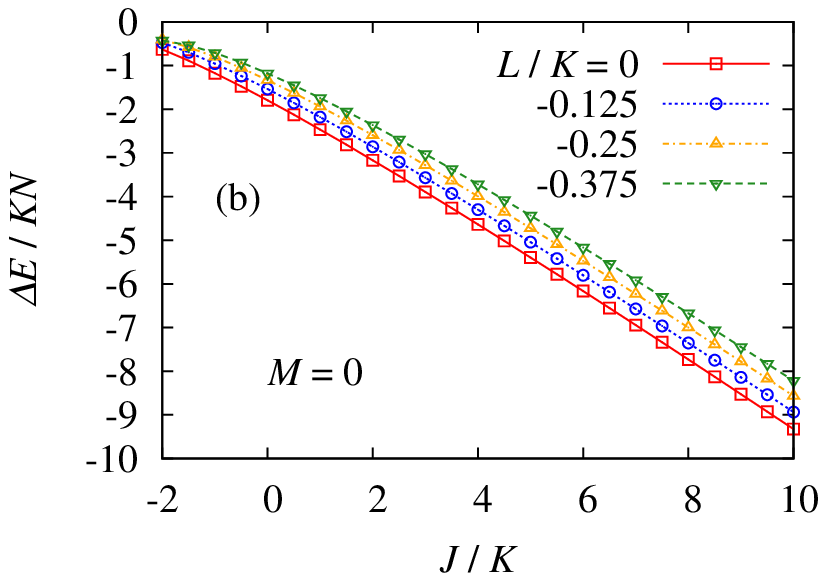}}
	\resizebox{0.45\textwidth}{!}{\includegraphics{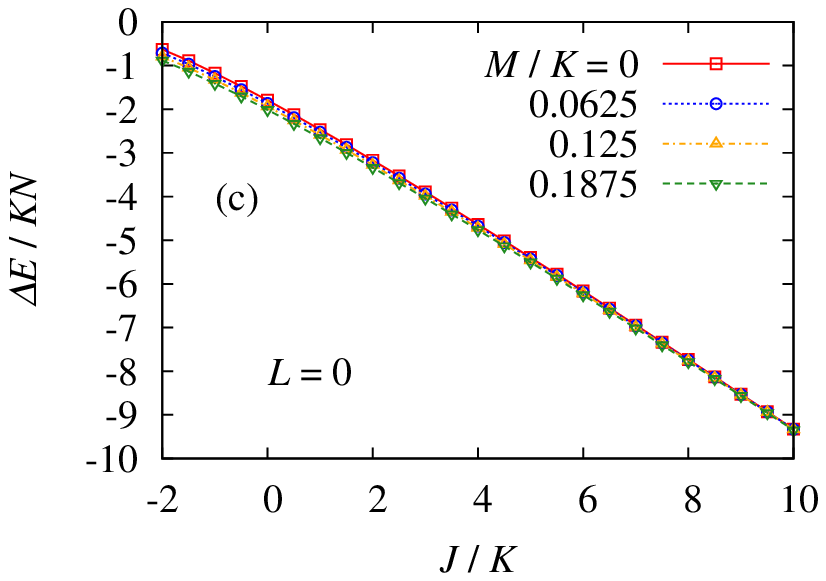}}
	\caption{$J / K$ dependences of quantum correction to the ground-state energy obtained on the basis of the spin-wave theory for (a) $M = -L / 2$, (b) $M = 0$, and (c) $L = 0$, with $J_{\rm 2nd}=J_{\rm 3rd}=0$.
	The lines are guides to the eyes.}
	\label{fig_crr_j5j6}
\end{figure}

The quantum correction of the ground-state energy is defined as
\begin{align}
\varDelta E
	= \frac{1}{2} \sum_{\mu = 1}^{4} \sum_{\mib{k}} \{ \omega_{\mu}^{}(\mib{k}) - X_{\mu}^{}(\mib{k}) \}~,
	\label{crr_j5j6}
\end{align}
which corresponds to the difference between the ground-state energy obtained on the basis of the spin-wave theory and the classical ground-state energy.
We show the $J / K$ dependences of the quantum correction per site for $M = -L / 2$ in Fig.~\ref{fig_crr_j5j6}(a). 
In addition, we show the results for $M = 0$ and $L = 0$ in Figs.~\ref{fig_crr_j5j6}(b) and \ref{fig_crr_j5j6}(c), respectively, to confirm the effect of each interaction.
The magnitude of $\varDelta E$ becomes small with decreasing $J / K$ for all parameter sets.
Namely, quantum mechanical effects are significant for a large positive $J / K$.
For example, $\varDelta E / KN \simeq -0.63$ and $-7.73$ at $J / K = -2$ and $8$, respectively, for $L = M = 0$.
The reduction in $\varDelta E / E^{\rm cl}$ is approximately 17.2\% for $J / K = -2$ but reaches approximately 56.6\% for $J / K = 8$, where $E^{\rm cl}$ is the ground-state energy in the classical system.
The reduction for the triangular Heisenberg antiferromagnet (THAF) is 43.9\%~\cite{Jolicoeur1989}.
For $M = -L / 2$ and $M = 0$, the magnitude of $\varDelta E$ decreases with increasing $| L | / K$. 
On the other hand, $\varDelta E$ hardly changes with $M/K$ in the $L = 0$ systems.

\subsubsection{Sublattice magnetization}

\begin{figure}[t]
	\centering
	\resizebox{0.45\textwidth}{!}{\includegraphics{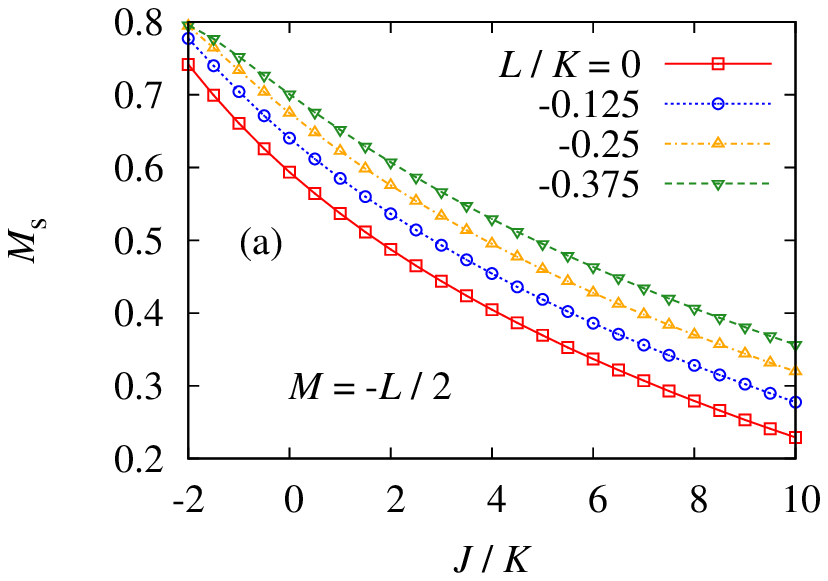}}\\
	\resizebox{0.45\textwidth}{!}{\includegraphics{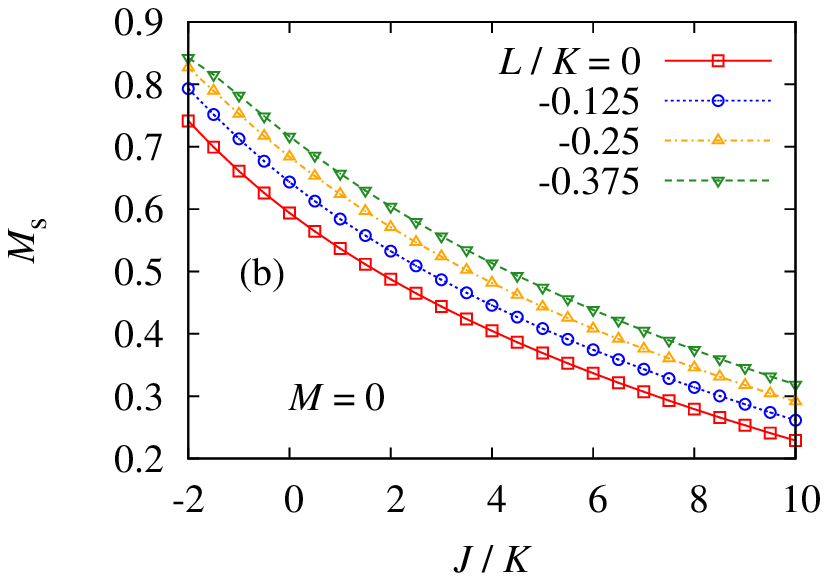}}
	\resizebox{0.45\textwidth}{!}{\includegraphics{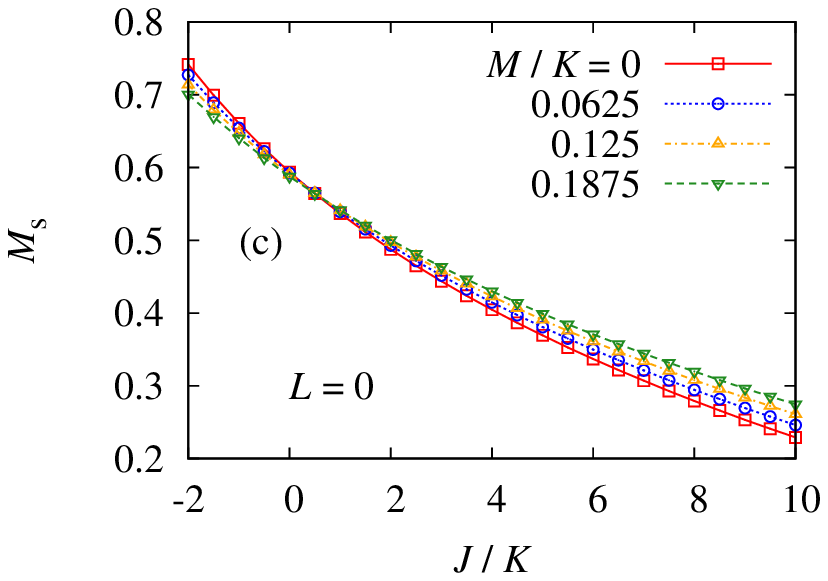}}
	\caption{$J / K$ dependences of the sublattice magnetizations obtained on the basis of the spin-wave theory for (a) $M = -L / 2$, (b) $M = 0$, and (c) $L = 0$ with $J_{\rm 2nd}=J_{\rm 3rd}=0$.
	The lines are guides to the eyes.}
	\label{mag_j5j6}
\end{figure}
The sublattice magnetization per site normalized by the classical value is given by
\begin{align}
M_{\rm s}
	= 1 - \frac{1}{N} \sum_{\mu = 1}^{4} \sum_{\mib{k}} \Bigg( \frac{X_{\mu}^{}(\mib{k})}{\omega_{\mu}^{}(\mib{k})} - 1 \Bigg)
	\label{mag_site}
\end{align}
in the ground state.
Although the magnetizations of the four sublattices have different directions, the magnitudes of the magnetizations are the same as $M_{\rm s}$ in the tetrahedral-structure state.
We show the $J / K$ dependences of $M_{\rm s}$ obtained on the basis of the spin-wave theory for three cases, $M = -L / 2$, $M = 0$, and $L = 0$, in ${\rm Figs.}~\ref{mag_j5j6}{\rm (a)}~\mbox{--}~\ref{mag_j5j6}{\rm (c)}$, respectively.
The value of $M_{\rm s}$ decreases with increasing $J / K$.
Namely, quantum mechanical effects are significant for a large positive $J / K$.
For example, $M_{\rm s} \simeq 0.74$ and $0.28$ at $J / K = -2$ and 8, respectively, for $L=M=0$.
The reduction is approximately 26\% for $J / K = -2$ but reaches approximately 72\% for $J / K = 8$.
The reduction for the THAF, in which numerical studies have suggested the existence of finite ground-state LRO~\cite{Singh1992, BernuPRL1992, Momoi1994}, is 52.2\% within the spin-wave theory~\cite{Jolicoeur1989, Momoi1992}.
In addition, the reduction for the triangular $XY$ antiferromagnet is estimated to be 56.3\% in the linear spin-wave theory~\cite{Leung1993} and about 59\% for finite-size systems~\cite{Momoi1994}.

Furthermore, Figs.~\ref{mag_j5j6}(a) and \ref{mag_j5j6}(b) show that $M_{\rm s}$ increases with $| L | / K$.
On the other hand, $M_{\rm s}$ hardly changes with $M/K$ in the $L = 0$ system in comparison with that in the $M=0$ system.

\subsubsection{Scalar chirality}

\begin{figure}[t]
	\centering
	\resizebox{0.45\textwidth}{!}{\includegraphics{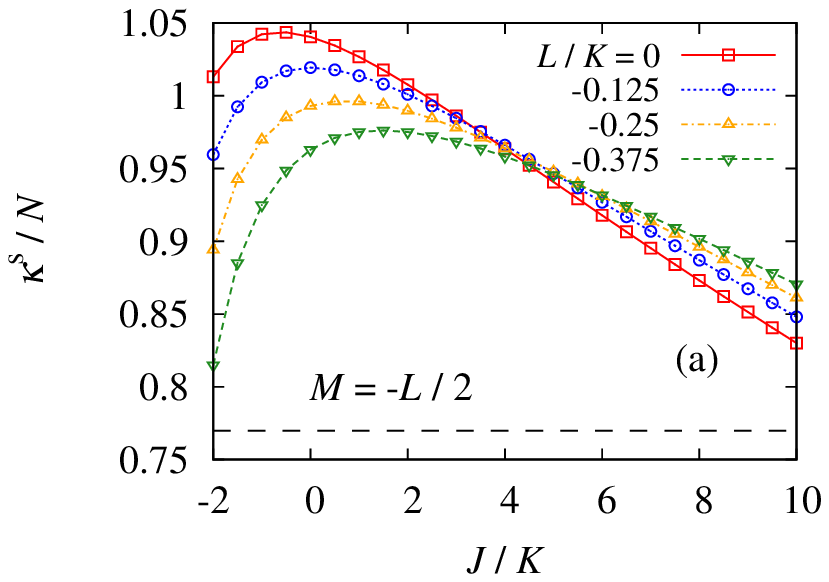}}\\
	\resizebox{0.45\textwidth}{!}{\includegraphics{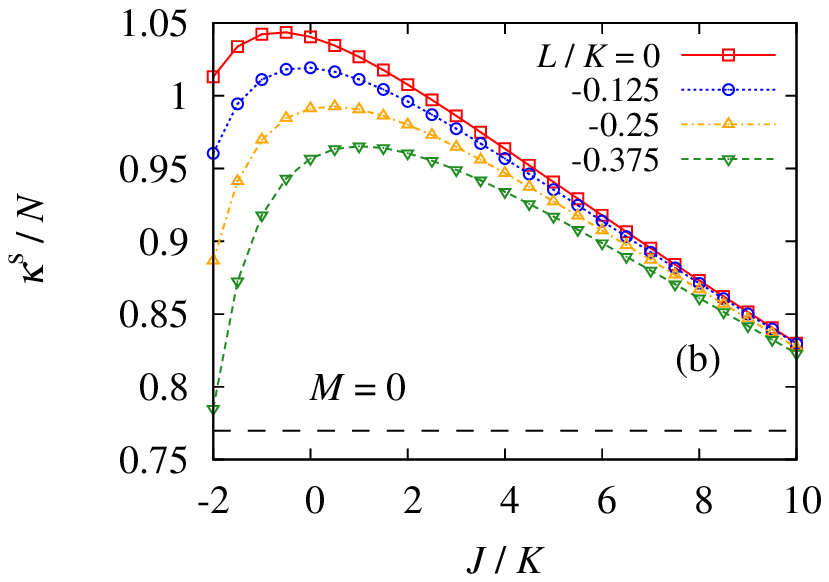}}
	\resizebox{0.45\textwidth}{!}{\includegraphics{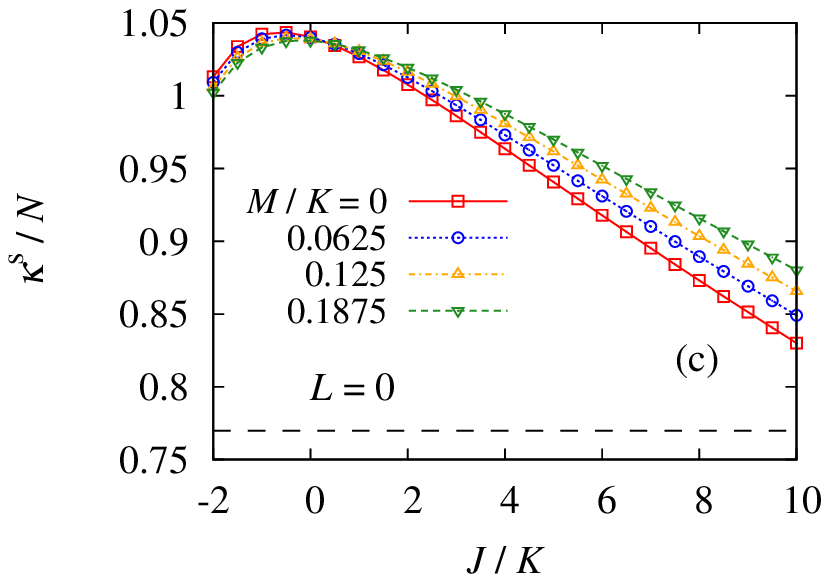}}
	\caption{$J / K$ dependences of the scalar chirality obtained on the basis of the spin-wave theory for (a) $M = -L / 2$, (b) $M = 0$, and (c) $L = 0$.
	The horizontal broken lines denote the classical value of the scalar chirality $\kappa^{\rm s} / N = 4 / 3\sqrt{3} \simeq 0.77$.
	The solid, dotted, chain, and broken lines are guides to the eyes.}
	\label{schiral_j5j6}
\end{figure}

The expectation value of the scalar chiral operator defined by Eq.~(\ref{schiral}) is written as
\begin{align}
\kappa^{\rm s}
	&\equiv \langle \hat{\kappa}^{\rm s} \rangle
	\notag \\
	&
	= \frac{4}{3\sqrt{3}} N
	- 4 \sum_{\mu = 1}^{4} \sum_{\mib{k}} \Big \{ F_{\mu}^{}(\mib{k}) + G_{\mu}^{}(\mib{k}) \Big \langle \alpha_{\mu,\mib{k}}^{\dag} \alpha_{\mu,\mib{k}}^{} \Big \rangle \Big \}~,
	\label{schiral_sw}
\end{align}
in terms of the spin-wave operators.
Explicit expressions for $F_{\mu}^{}(\mib{k})$ and $G_{\mu}^{}(\mib{k})$ are given in the Appendix.
Because $\Big \langle \alpha_{\mu,\mib{k}}^{\dag} \alpha_{\mu,\mib{k}}^{} \Big \rangle = 0$ in the ground state, the scalar chirality per upward-pointing triangle in the ground state is described by
\begin{align}
\frac{\kappa^{\rm s}}{N}
	= \frac{4}{3\sqrt{3}} - \frac{4}{N} \sum_{\mu = 1}^{4} \sum_{\mib{k}} F_{\mu}^{}(\mib{k})~.
	\label{schiral_ground_state}
\end{align}
We show the $J / K$ dependences of the scalar chirality obtained on the basis of the spin-wave theory for three cases, $M = -L / 2$, $M = 0$, and $L = 0$, in ${\rm Figs.}~\ref{schiral_j5j6}{\rm (a)}~\mbox{--}~\ref{schiral_j5j6}{\rm (c)}$, respectively.
The horizontal broken lines in Fig.~\ref{schiral_j5j6} denote the classical value of the scalar chirality $\kappa^{\rm s} / N = 4 / 3\sqrt{3} \simeq 0.77$.
The values of $\kappa^{\rm s}$ are larger than the classical value in the parameter sets that we calculated.
In the quantum model, the largest eigenvalue of $\mib{\sigma}_{1}^{} \cdot (\mib{\sigma}_{2}^{} \times \mib{\sigma}_{3}^{})$ on a triangle is larger than the classical value, unlike the sublattice magnetization. 
Therefore, the scalar chirality of the quantum system may also be larger than the classical value.
A similar tendency appears for the vector chirality in the triangular $XXZ$ antiferromagnet.~\cite{Momoi1992}
Note that a $J/K$ dependence of $\kappa^{\rm s}$ for $L=M=0$ was reported in previous studies~\cite{Momoi1997, Kubo1998-2} but is not consistent with our result.
We have thus corrected the previous result.

It is difficult to interpret the effects of quantum fluctuations on the scalar chirality in comparison with the energy and sublattice magnetization because the values of $\kappa^{\rm s}$ exceed the classical one.
From the results of $\varDelta E$ and $M_{\rm s}$, it is reasonable to examine the effects of quantum fluctuations by using the increase or decrease in $\kappa^{\rm s}$ itself rather than by comparison with the classical value.
A similar $J/K$ dependence of $\kappa^{\rm s}$ is seen for all parameter sets, as shown in Fig.~\ref{schiral_j5j6}.
Specifically, a peak for $\kappa^{\rm s}$ exists at the midpoint of $J/K$ and the value of $\kappa^{\rm s}$ decreases as $J/K$ approaches $-2$ or $10$.
The decrease in $\kappa^{\rm s}$ in the antiferromagnetic $J/K$ region shows that quantum mechanical effects are significant for a large positive $J / K$,
as seen from the results of $\varDelta E$ and $M_{\rm s}$.
On the other hand, the decrease in $\kappa^{\rm s}$ in the ferromagnetic $J / K$ region, which is not seen for $\varDelta E$ and $M_{\rm s}$, might be caused by the instability of the tetrahedral structure owing to the approach to the softening point outside the tetrahedral phase in the classical system.

Figure~\ref{schiral_j5j6}(b) shows that $\kappa^{\rm s}$ decreases with increasing $|L|/K$ for $M=0$ and that the trend is noticeable as $J/K$ approaches $-2$.
This result can be understood from the ferromagnetic five-spin interactions cooperating with the ferromagnetic $J/K$.
On the other hand, Fig.~\ref{schiral_j5j6}(c) shows that $\kappa^{\rm s}$ increases with increasing $M/K$ for $L=0$ and that the trend is noticeable as $J/K$ approaches $10$.
The result shows that the six-spin interactions stabilize the tetrahedral structure in the quantum system.
The dependences of $\kappa^{\rm s}$ on $J/K$, $L/K$, and $M/K$ for $M=-L/2$ shown in Fig.~\ref{schiral_j5j6}(a) also show similar trends to those seen in Figs.~\ref{schiral_j5j6}(b) and \ref{schiral_j5j6}(c).

\subsection{$J$-$K$-$J_{\rm 2nd}^{}$-$J_{\rm 3rd}^{}$ model}

Next, we set the parameters $L = M = 0$ and describe the physical quantities in the $J$-$K$-$J_{\rm 2nd}^{}$-$J_{\rm 3rd}^{}$ model.
In the previous subsection, we explained that the effects of quantum fluctuations on the physical quantities $\Delta E$ and $M_{\rm s}$ are weak for large $|L| / K$.
In the $J$-$K$-$J_{\rm 2nd}^{}$-$J_{\rm 3rd}^{}$ model, the effects of quantum fluctuations on the physical quantities are weak for the antiferromagnetic second-nearest-neighbor and the ferromagnetic third-nearest-neighbor interactions.
The details of the results in the $J$-$K$-$J_{\rm 2nd}^{}$-$J_{\rm 3rd}^{}$ model will be explained below.

\subsubsection{Quantum correction of the ground-state energy}

\begin{figure}[t]
	\centering
	\resizebox{0.45\textwidth}{!}{\includegraphics{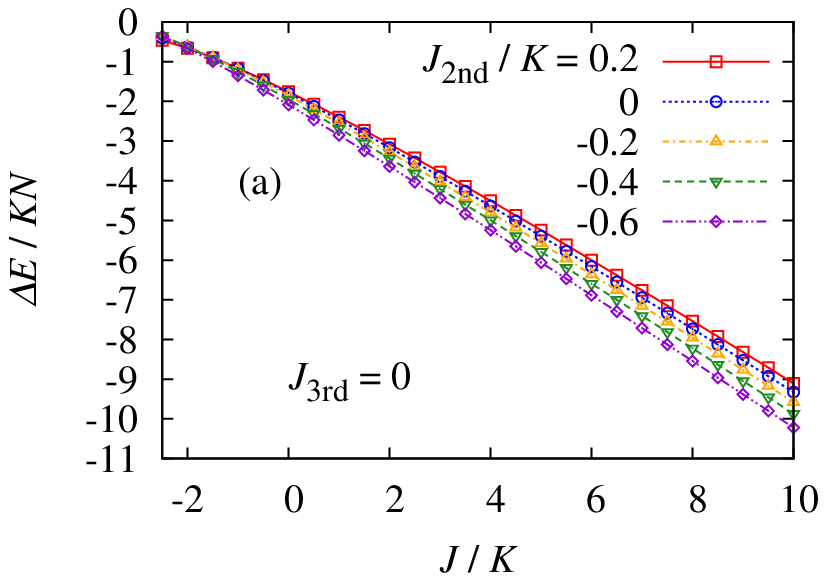}}
	\resizebox{0.45\textwidth}{!}{\includegraphics{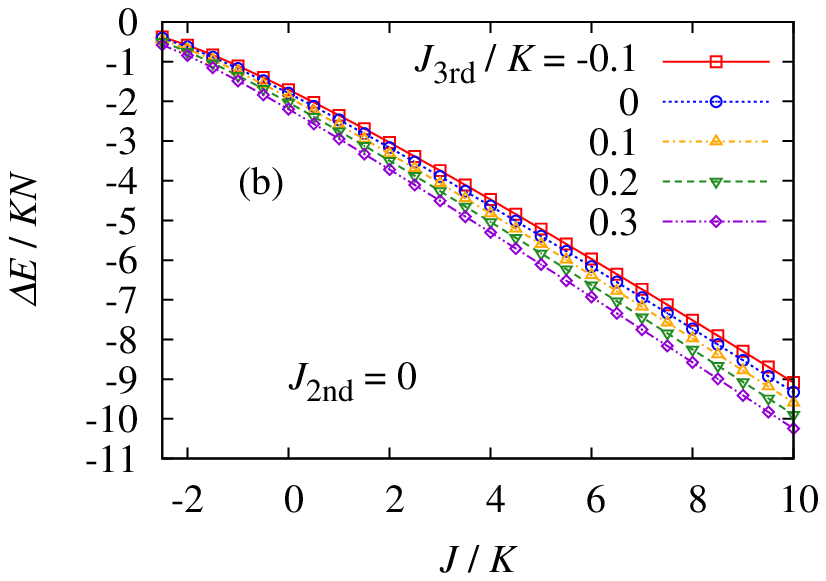}}
	\caption{$J / K$ dependences of the quantum correction to the ground-state energy per site obtained on the basis of the spin-wave theory for (a) $J_{\rm 3rd}^{} = 0$ and (b) $J_{\rm 2nd}^{} = 0$ with $L = M = 0$.
	The lines are guides to the eyes.}
	\label{fig_ene_crr_j2j3}
\end{figure}

We show the $J / K$ dependences of the quantum correction to the ground-state energy per site obtained on the basis of the spin-wave theory for the cases $J_{\rm 3rd}^{} = 0$ and $J_{\rm 2nd}^{} = 0$ in Figs.~\ref{fig_ene_crr_j2j3}(a) and \ref{fig_ene_crr_j2j3}(b), respectively.
The magnitude of $\varDelta E$ increases with increasing $J / K$ for all parameter sets in both the $J_{\rm 3rd}^{} = 0$ and $J_{\rm 2nd}^{} = 0$ systems.
Therefore, quantum mechanical effects are significant for a large positive $J / K$.
The magnitude of $\varDelta E$ also increases with decreasing $J_{\rm 2nd}^{} / K$ at $J_{\rm 3rd}^{} = 0$ and with increasing $J_{\rm 3rd}^{} / K$ at $J_{\rm 2nd}^{} = 0$ for all parameter sets.

\subsubsection{Sublattice magnetization}

\begin{figure}[t]
	\centering
	\resizebox{0.45\textwidth}{!}{\includegraphics{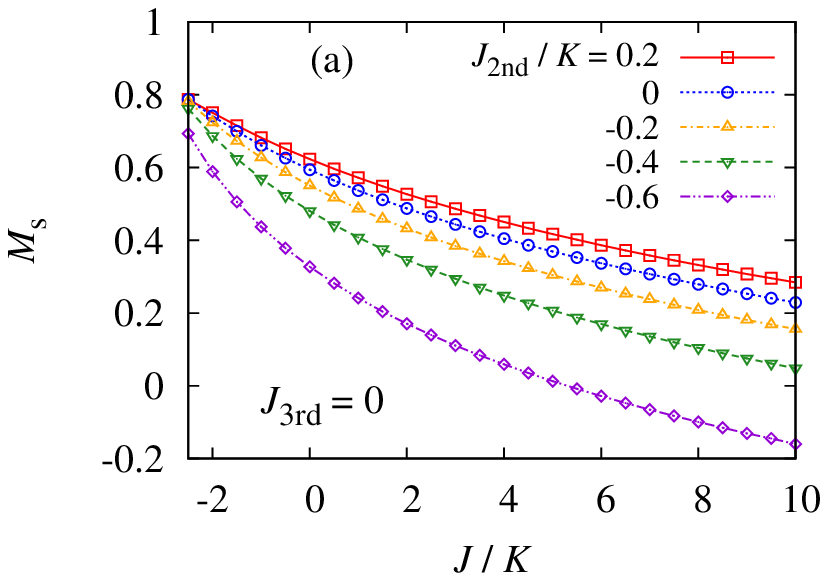}}
	\resizebox{0.45\textwidth}{!}{\includegraphics{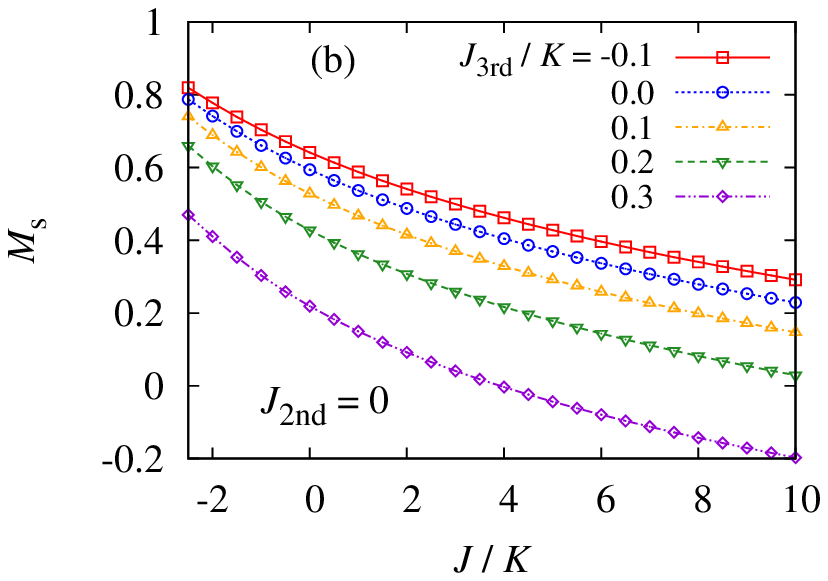}}
	\caption{$J / K$ dependences of the sublattice magnetization obtained on the basis of the spin-wave theory for (a) $J_{\rm 3rd}^{} = 0$ and (b) $J_{\rm 2nd}^{} = 0$ with $L = M = 0$.
	The lines are guides to the eyes.}
	\label{fig_mag_j2j3}
\end{figure}

We show the $J / K$ dependences of the sublattice magnetization obtained on the basis of the spin-wave theory for the cases $J_{\rm 3rd}^{} = 0$ and $J_{\rm 2nd}^{} = 0$ in Figs.~\ref{fig_mag_j2j3}(a) and \ref{fig_mag_j2j3}(b), respectively.
The value of $M_{\rm s}$ decreases with increasing $J / K$ for all parameter sets in both the $J_{\rm 3rd}^{} = 0$ and $J_{\rm 2nd}^{} = 0$ systems.
This result shows that quantum mechanical effects are significant for a large positive $J / K$.
The value of $M_{\rm s}$ also decreases with decreasing $J_{\rm 2nd}^{} / K$ at $J_{\rm 3rd}^{} = 0$ and with increasing $J_{\rm 3rd}^{} / K$ at $J_{\rm 2nd}^{} = 0$.
These tendencies are the same as those of the quantum correction to the ground-state energy.

\subsubsection{Scalar chirality}

\begin{figure}[t]
	\centering
	\resizebox{0.45\textwidth}{!}{\includegraphics{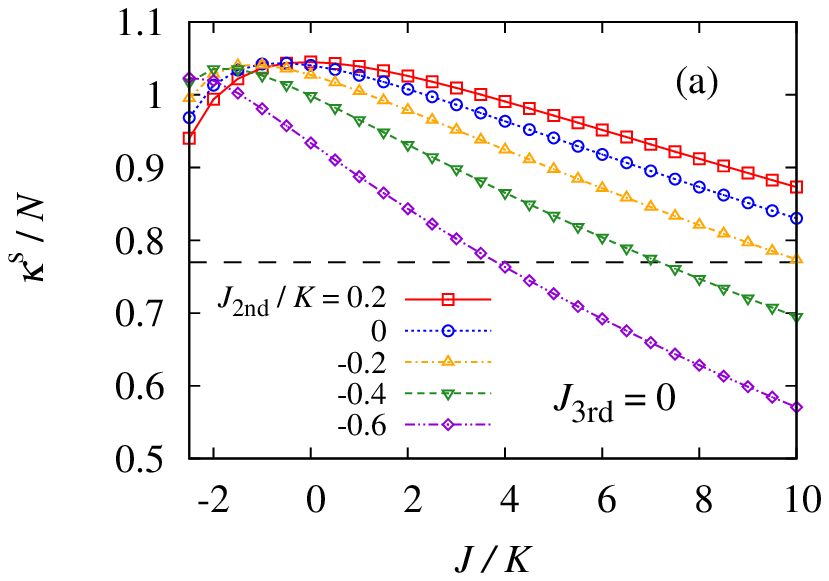}}
	\resizebox{0.45\textwidth}{!}{\includegraphics{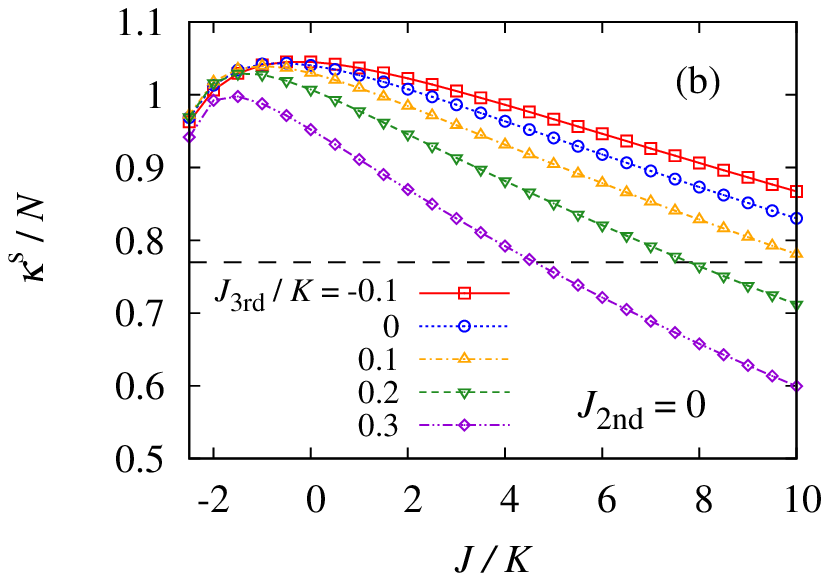}}
	\caption{$J / K$ dependences of the scalar chirality obtained on the basis of the spin-wave theory for (a) $J_{\rm 3rd}^{} = 0$ and (b) $J_{\rm 2nd}^{} = 0$ with $L = M = 0$.
	The horizontal broken lines denote the classical value $\kappa^{\rm s} / N = 4 / 3\sqrt{3} \simeq 0.77$.
	The lines, except for the horizontal broken line, are guides to the eyes.}
	\label{fig_schiral_j2j3}
\end{figure}
We show the $J / K$ dependences of the scalar chirality obtained on the basis of the spin-wave theory for the cases $J_{\rm 3rd}^{} = 0$ and $J_{\rm 2nd}^{} = 0$ in Figs.~\ref{fig_schiral_j2j3}(a) and \ref{fig_schiral_j2j3}(b), respectively.
The horizontal broken lines in Fig.~\ref{fig_schiral_j2j3} denote the classical value $\kappa^{\rm s} / N = 4 / 3\sqrt{3} \simeq 0.77$.
There are regions of $J / K$ where the quantum values of $\kappa^{\rm s}$ are larger than the classical value.
The effects of quantum fluctuations are the same as those on the energy correction and sublattice magnetization, except for a large negative $J / K$.
Namely, quantum mechanical effects become significant with both decreasing $J_{\rm 2nd}^{} / K$ for $J / K \geq 0$ in the $J_{\rm 3rd}^{} = 0$ system and with increasing $J_{\rm 3rd}^{} / K$ for $J / K \geq -0.5$ in the $J_{\rm 2nd}^{} = 0$ system.
The decreases for large negative $J / K$ might be caused by the instability of the tetrahedral structure owing to the approach to the softening point outside the tetrahedral phase in the classical system as mentioned in Sect.~5.1.


\section{Summary and Discussion}

In this paper, we have studied the effects of quantum fluctuations on the tetrahedral structure in the ground state of the MSE model with up to the six-spin exchange interactions on a triangular lattice using the linear spin-wave theory.
First, we determined the region of $J/K$ in which the tetrahedral structure is stable as the ground state within the mean-field approximation, assuming $6 \times 6$ and $12 \times 12$ sublattices.
While the region shrinks for large five-spin interactions, the region hardly depends on the six-spin interactions.
We also investigated the effects of the second- and third-nearest-neighbor interactions on the ground state instead of the five- and six-spin interactions.
For $J_{\rm 3rd}^{} = 0$, the region of $J/K$ in which the ground state is the tetrahedral-structure state expands with $J_{\rm 2nd}^{}$ for $J_{\rm 2nd}^{} > 0$.
On the other hand, for $J_{\rm 2nd}^{} < 0$, the region shrinks with increasing $|J_{\rm 2nd}^{}|$.
When the value of $J_{\rm 3rd}^{}$ changes for $J_{\rm 2nd}^{} = 0$, we obtained the opposite trend to $J_{\rm 2nd}^{}$.
Because two sites in the tetrahedral structure connected by $J_{\rm 2nd}^{}$ and $J_{\rm 3rd}^{}$ always belong to different and the same sublattices, respectively, it is easy to understand these results.

Next, we applied the linear spin-wave theory in the parameter region where the tetrahedral structure is realized as the classical ground state.
We concluded that the tetrahedral structure survives the quantum fluctuations within the linear spin-wave theory because no softening of the spin wave was induced in the whole region of the tetrahedral-structure phase. 
Calculating the quantum corrections to the ground-state energy and the sublattice magnetization, we found that the effects of quantum fluctuations are weak for small $J/K$ and large $|L|/K$, where the system approaches the ferromagnetic phase.
Furthermore, for antiferromagnetic $J_{\rm 2nd}^{}$ and ferromagnetic $J_{\rm 3rd}^{}$, the effects of the quantum fluctuations are weak.
On the other hand, the six-spin interaction $M/K$ has little effect on the quantum corrections.
The results show that the tetrahedral structure is robust against $M/K$ not only within the mean-field approximation but also in the spin-wave theory.

It is difficult to interpret the effects of quantum fluctuations on the scalar chirality in comparison with the energy and sublattice magnetization because the scalar chirality exceeds the classical one.
We gave the correct $J/K$ dependence of the scalar chirality, which disagrees with that reported in previous studies~\cite{Momoi1997, Kubo1998-2} for the multiple-spin exchange model with up to the four-spin exchange interactions.
A decrease in the scalar chirality, which was not seen for the energy and sublattice magnetization, was seen for small $J/K$ and large $|L|/K$.
This result might be caused by the instability of the tetrahedral structure owing to the approach to the softening point outside the tetrahedral phase in the classical system.
Furthermore, the scalar chirality shows that the six-spin interactions and antiferromagnetic second- and ferromagnetic third-nearest-neighbor interactions stabilize the tetrahedral structure in the quantum system.

We thus find that, in the linear spin-wave theory, the tetrahedral-structure state survives the quantum fluctuations.
However, the exact-diagonalization results of the MSE model with up to the six-spin interactions suggest that the ground state in the parameter region where the tetrahedral structure occurs within the mean-field approximation is in a nonmagnetic gapped spin-liquid phase~\cite{Misguich1998, Misguich1999}.
Furthermore, the exact-diagonalization study of the MSE model with up to the four-spin interactions suggests that a gapless spin-liquid state exists on the antiferromagnetic side of the tetrahedral-structure phase for the classical system~\cite{LiMing2000}.
If the ground state of the quantum systems is the quantum spin-liquid state, it would be interesting to study the quantum mechanism that destabilizes the tetrahedral-structure state.
One of the disadvantages in either the linear spin-wave theory or the exact-diagonalization study is considered as a cause of this discrepancy in the results between the methods.
One of the possible disadvantages is the inadequacy of the linear approximation in the spin-wave expansion, and the other is the smallness of the system size in the exact-diagonalization study.
So far, the cause of the discrepancy has not been clarified yet and solving this issue remains a future problem.
The effects of higher-order terms in the spin-wave expansion have been studied on the N${\rm {\acute e}}$el phase in square-lattice systems with the four-spin interactions~\cite{Katanin2002, Majumdar2012}.
Future studies on the effects of higher-order terms in the present system are desired.


\begin{acknowledgments}
The authors acknowledge stimulating discussions with Philippe Sindzingre.
This work was supported by JSPS KAKENHI Grant Numbers JP16K05425 and JP16K05479.
\end{acknowledgments}


\appendix

\section{Calculation of the Scalar Chirality}

From the Holstein--Primakoff and unitary transformations, the scalar chirality described by Eq.~(\ref{schiral}) is written as
\begin{align}
\hat{\kappa}^{\rm s}
	&= \frac{4}{3\sqrt{3}} N
	- \frac{8}{3\sqrt{3}} \sum_{\mu = 1}^{4} {\sum_{\mib{k}}}' \Big \{
	f_{\mu}^{}(\mib{k}) \Big( {\tilde a}_{\mu,\mib{k}}^{\dag}{\tilde a}_{\mu,\mib{k}}^{} + {\tilde a}_{\mu,-\mib{k}}^{\dag}{\tilde a}_{\mu,-\mib{k}}^{} \Big)
	\notag \\
	&
	+ y_{\mu}^{}(\mib{k}) {\tilde a}_{\mu,-\mib{k}}^{}{\tilde a}_{\mu,\mib{k}}^{}
	+ y_{\mu}^{\ast}(\mib{k}) {\tilde a}_{\mu,-\mib{k}}^{\dag}{\tilde a}_{\mu,\mib{k}}^{\dag}\Big \}~,
	\label{schiral_unitary}
\end{align}
where
\begin{align}
f_{1}^{}(\mib{k})
	&= 3 + \cos k_{2}^{} + \cos (k_{1}^{} - k_{2}^{}) + \cos k_{1}^{}~,
	\notag \\
f_{2}^{}(\mib{k})
	&= 3 + \cos k_{2}^{} - \cos (k_{1}^{} - k_{2}^{}) - \cos k_{1}^{}~,
	\notag \\
f_{3}^{}(\mib{k})
	&= 3 - \cos k_{2}^{} + \cos (k_{1}^{} - k_{2}^{}) - \cos k_{1}^{}~,
	\notag \\
f_{4}^{}(\mib{k})
	&= 3 - \cos k_{2}^{} - \cos (k_{1}^{} - k_{2}^{}) + \cos k_{1}^{}~,
	\notag \\
y_{1}^{}(\mib{k})
	&= \cos k_{2}^{} + \phi^{\ast}\cos (k_{1}^{} - k_{2}^{}) + \phi \cos k_{1}^{}~,
	\notag \\
y_{2}^{}(\mib{k})
	&= \cos k_{2}^{} - \phi^{\ast}\cos (k_{1}^{} - k_{2}^{}) - \phi \cos k_{1}^{}~,
	\notag \\
y_{3}^{}(\mib{k})
	&= -\cos k_{2}^{} + \phi^{\ast}\cos (k_{1}^{} - k_{2}^{}) - \phi \cos k_{1}^{}~,
	\notag \\
y_{4}^{}(\mib{k})
	&= -\cos k_{2}^{} - \phi^{\ast}\cos (k_{1}^{} - k_{2}^{}) + \phi \cos k_{1}^{}~.
	\label{coef_f_y}
\end{align}
Furthermore, performing the Bogoliubov transformation and taking the average, we obtain the expectation value of the scalar chirality
\begin{align}
\kappa^{\rm s}
	= \frac{4}{3\sqrt{3}} N
	- 4 \sum_{\mu = 1}^{4} \sum_{\mib{k}} \Big \{ F_{\mu}^{}(\mib{k}) + G_{\mu}^{}(\mib{k}) \Big \langle \alpha_{\mu,\mib{k}}^{\dag} \alpha_{\mu,\mib{k}}^{} \Big \rangle \Big \}~,
	\label{schiral_Bogoliubov}
\end{align}
where
\begin{align}
F_{\mu}^{}(\mib{k})
	&= \frac{1}{2} G_{\mu}^{}(\mib{k}) - \frac{1}{3\sqrt{3}} f_{\mu}^{}(\mib{k})~,
	\notag \\
G_{\mu}^{}(\mib{k})
	&= \frac{2}{3\sqrt{3} \omega_{\mu}^{}(\mib{k})} \{ f_{\mu}^{}(\mib{k}) X_{\mu}^{}(\mib{k}) - g_{\mu}^{}(\mib{k})\}~,
	\notag \\
g_{1}^{}(\mib{k})
	&= \frac{1}{2} [\{ 2\cos k_{2}^{} - \cos (k_{1}^{} - k_{2^{}}) - \cos k_{1}^{} \} C_{1}^{}(\mib{k})
	\notag \\
	&~~~~~
	~~
	+ \{ -\cos k_{2}^{} + 2\cos (k_{1}^{} - k_{2^{}}) - \cos k_{1}^{} \} C_{2}^{}(\mib{k})
	\notag \\
	&~~~~~
	~~
	+ \{ -\cos k_{2}^{} - \cos (k_{1}^{} - k_{2^{}}) + 2\cos k_{1}^{} \} C_{3}^{}(\mib{k})]~,
	\notag \\
g_{2}^{}(\mib{k})
	&= \frac{1}{2} [\{ 2\cos k_{2}^{} + \cos (k_{1}^{} + k_{2^{}}) + \cos k_{1}^{} \} C_{1}^{}(\mib{k})
	\notag \\
	&~~~~~
	~~
	+ \{ \cos k_{2}^{} + 2\cos (k_{1}^{} - k_{2^{}}) - \cos k_{1}^{} \} C_{2}^{}(\mib{k})
	\notag \\
	&~~~~~
	~~
	+ \{ \cos k_{2}^{} - \cos (k_{1}^{} - k_{2^{}}) + 2\cos k_{1}^{} \} C_{3}^{}(\mib{k})]~,
	\notag \\
g_{3}^{}(\mib{k})
	&= \frac{1}{2} [\{ 2\cos k_{2}^{} + \cos (k_{1}^{} + k_{2^{}}) - \cos k_{1}^{} \} C_{1}^{}(\mib{k})
	\notag \\
	&~~~~~
	~~
	+ \{ \cos k_{2}^{} + 2\cos (k_{1}^{} - k_{2^{}}) + \cos k_{1}^{} \} C_{2}^{}(\mib{k})
	\notag \\
	&~~~~~
	~~
	+ \{ -\cos k_{2}^{} + \cos (k_{1}^{} - k_{2^{}}) + 2\cos k_{1}^{} \} C_{3}^{}(\mib{k})]~,
	\notag \\
g_{4}^{}(\mib{k})
	&= \frac{1}{2} [\{ 2\cos k_{2}^{} - \cos (k_{1}^{} + k_{2^{}}) + \cos k_{1}^{} \} C_{1}^{}(\mib{k})
	\notag \\
	&~~~~~
	~~
	+ \{ -\cos k_{2}^{} + 2\cos (k_{1}^{} - k_{2^{}}) + \cos k_{1}^{} \} C_{2}^{}(\mib{k})
	\notag \\
	&~~~~~
	~~
	+ \{ \cos k_{2}^{} + \cos (k_{1}^{} - k_{2^{}}) + 2\cos k_{1}^{} \} C_{3}^{}(\mib{k})]~.
	\label{coef_F_G_g}
\end{align}

We examine the behavior of $F_{\mu}^{}(\mib{k})$ and $G_{\mu}^{}(\mib{k})$ for $\mib{k} \simeq \mib{0}$.
For $\mu = 1$, we obtain $f_{1}^{} (\mib{0}) = 6$, $g_{1}^{}(\mib{0}) = 0$, $G_{1}^{}(\mib{0}) = 4 / \sqrt{3}$, and $F_{1}^{}(\mib{0}) = 0$.
Because $f_{1}^{} (\mib{k})$, $f_{1}^{} (\mib{k}) X_{1}^{}(\mib{k}) - g_{1}^{}(\mib{k})$, and $\omega_{1}^{}(\mib{k})$ are proportional to $k^{2}$, 
$G_{1}^{}(\mib{k})$ and $F_{1}^{}(\mib{k})$ are proportional to $k^{0}$ and $k^{2}$, respectively.
For $\mu = 2,3,4$, we obtain $f_{\mu}^{} (\mib{0}) = 2$, $g_{1}^{}(\mib{0}) = 32(9J + 9J_{\rm 2nd}^{} + 36K + 24L + 32M) / 27$, $G_{\mu}^{}(\mib{0}) = 0$, and $F_{\mu}^{}(\mib{0}) = -2 / 3\sqrt{3}$.
Because $f_{\mu}^{} (\mib{k}) X_{\mu}^{}(\mib{k}) - g_{\mu}^{}(\mib{k})$ and $\omega_{\mu}^{}(\mib{k})$ are proportional to $k^{2}$ and $k$, $G_{\mu}^{}(\mib{k})$ and $F_{\mu}^{}(\mib{k})$ are proportional to $k$ and $k^{2}$, respectively.




\end{document}